\documentclass[11pt,a4paper]{article}

\usepackage[utf8]{inputenc}   % Encoding
\usepackage[T1]{fontenc}
\usepackage{lmodern}
\usepackage{amsmath,amssymb,amsthm}
\usepackage{graphicx}
\usepackage{hyperref}
\usepackage{geometry}
\usepackage{authblk}          % For multiple authors/affiliations
\usepackage{setspace}         % For line spacing
\usepackage[gen]{eurosym}     % For Euro symbol
\usepackage{siunitx}          % For scientific notation
\usepackage{authblk}

\providecommand{\keywords}[1]
{%
    \small \textbf{Keywords:} #1
}

\title{\textbf{Non-parametric Causal Discovery for EU Allowances Returns Through the Information Imbalance}}

\author[1]{Cristiano Salvagnin\thanks{Corresponding author: \texttt{cristiano.salvagnin@uninsubria.it}}}
\author[2]{Vittorio del Tatto}
\author[3]{Maria Elena De Giuli}
\author[4]{Antonietta Mira\thanks{Corresponding author: \texttt{antonietta.mira@usi.ch}}}
\author[5]{Aldo Glielmo\thanks{Corresponding author: \texttt{Aldo.Glielmo@bancaditalia.it}}}

\affil[1*]{Department of Science and High Technology, University of Insubria, Via Valleggio 11, 22100, Como, Italy.}
\affil[2]{Physics Section, Scuola Internazionale Superiore di Studi Avanzati (SISSA), Via Bonomea 265, 34136, Trieste, Italy.}
\affil[3]{Department of Economics and Management, University of Pavia, Via S. Felice Al Monastero 5, 27100, Pavia, Italy.}
\affil[4*]{Faculty of Economics, Euler Institute, Università della Svizzera Italiana (USI), Via Giuseppe Buffi 13, 6900, Lugano, Switzerland.}
\affil[4]{Department of Science and High Technology, University of Insubria, Via Valleggio 11, 22100, Como, Italy.}
\affil[6*]{Applied Research Team, Directorate General for IT, Banca d'Italia, Largo Guido Carli 1, 00044, Frascati (RM), Italy.}

\date{}

\begin{document}

\maketitle

% ===== ABSTRACT =====
\begin{abstract}
We propose to use a recently introduced non-parametric tool named Differentiable Information Imbalance (DII) to identify variables that are causally related -- potentially through non-linear relationships -- to the financial returns of the European Union Allowances (EUAs) within the EU Emissions Trading System (EU ETS). We examine data from January 2013 to April 2024 and compare the DII approach with multivariate Granger causality, a well-known linear approach based on VAR models. We find significant overlap among the causal variables identified by linear and non-linear methods, such as the coal futures prices and the IBEX35 index. We also find important differences between the two causal sets identified. On two synthetic datasets, we show how these differences could originate from limitations of the linear methodology.
\end{abstract}

% ===== Keywords =====
\newpage
% Section 1: Introduction -----------------------------------
\section{Introduction}
\label{sec1:Introduction}
\noindent
    The European Union Emission Trading System (EU ETS) is a key element in the EU's strategy to address climate change and reduce greenhouse gas (GHG) emissions. 
    Based on the cap-and-trade principle, this system establishes a gradual limit on GHG emissions across important sectors of the economy, primarily targeting energy, aviation, and energy-intensive industry.
    Under the EU ETS, participating companies receive emission permits, known as European Union Allowances (EUA), each granting the release of one tonne of carbon dioxide or its equivalent into the atmosphere.

\begin{figure}[htp]
        \centering
        \includegraphics[width=0.9\textwidth]{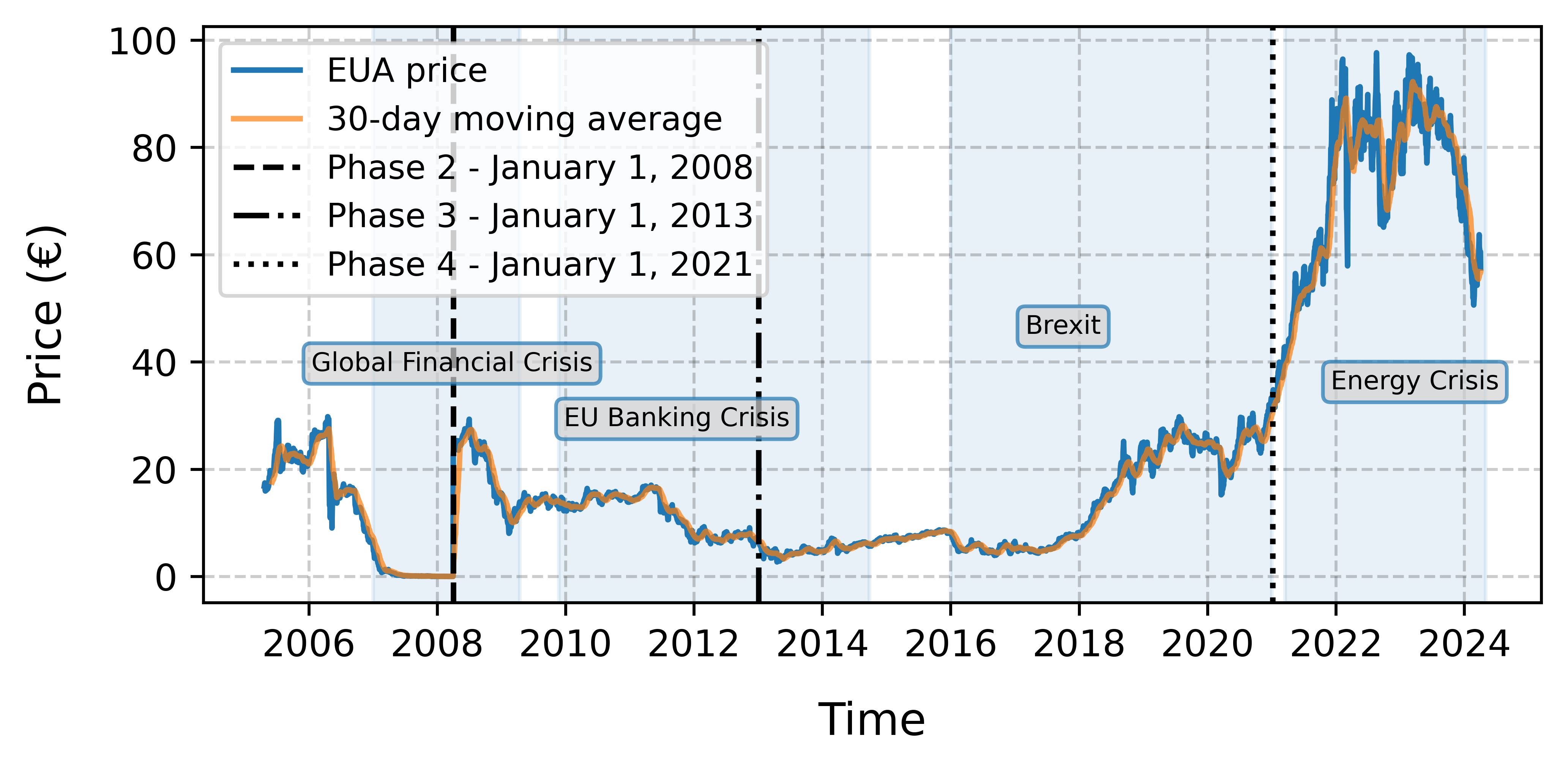}
        \vspace{-0.3cm}
        \caption{\textbf{EUA futures price.}
            The EUA prices analysed in this study represent futures contracts traded on the ICE Futures Europe, a major exchange for European carbon allowances. The figure shows EUA price from April 2005 to April 2024. Each market phase is marked by dashed vertical lines: Phase 1 (2005-2007), Phase 2 (2008-2012), Phase 3 (2013-2020), and Phase 4 (2020-present). Significant European economic and financial crises that have influenced EUA pricing are highlighted in light blue.}
        \label{fig:1}
    \end{figure}

    Companies participating in the market have the flexibility to trade these emission allowances on the ETS market.
    If a company exceeds its allocated allowances it must purchase additional permits to cover excess emissions, otherwise heavy fines, \euro100 per tonne of GHG emissions, are faced \cite{euets}.
    This market-driven approach provides an economic instrument for addressing GHG emissions efficiently. 
    Companies are encouraged to invest in emission reduction initiatives and technology to lower their emissions below their allocated allowances, enabling them to sell surplus permits for profit.
    Furthermore, the EU ETS undergoes regular reviews and adjustments to emission limits, with a key mechanism being the Linear Reduction Factor~(LRF).
    The LRF annually reduces the cap, ensuring that emission reduction targets remain consistently aligned with the EU's broader climate goals and evolving environmental priorities.
    Through the LRF, the market can be driven towards progressively more ambitious emission reduction targets while encouraging innovation and investment in cleaner technologies. 
    In essence, the EU ETS is a sophisticated market-based mechanism that leverages economic incentives to drive emissions reductions. 
    Its alignment with international agreements, such as the Kyoto Protocol (2005) and the Paris Agreement (2016), further highlights its importance as a critical tool in the global effort to combat climate change. The historical evolution of EUA futures prices and the phases of the EU ETS are illustrated in Figure~\ref{fig:1}.

%
% Subsection: Literature Review --------------------------------------
    \subsection{Literature review}
    \label{ssec1.1:LiteratureReview}

    \noindent
%% Intro
        Understanding causal relationships in financial time series is essential for interpreting market behaviour, forecasting asset dynamics, and measuring volatility effects. In the context of increasingly complex and interconnected financial systems, particularly those involving energy and commodity markets, accurate identification of causality goes beyond mere correlation and requires robust analytical tools.
        
        Over the years, a wide range of methods have been developed for this purpose, beginning with linear models such as Granger causality and Vector Autoregressions (VARs), which remain influential due to their simplicity and interpretability. However, the limitations of these traditional approaches in capturing non-linear dependencies have prompted the emergence of alternative techniques, including non-parametric measures like Transfer Entropy (TE)
        %, machine learning models such as Long Short-Term Memory (LSTM) networks and Random Forests, 
        and recent advances in information-theoretic metrics.
        This literature review provides an overview of these methodologies, highlighting their respective strengths and weaknesses, and positioning the Differentiable Information Imbalance (DII) metric within this evolving landscape of causal discovery tools.

%%  Traditional Linear Approaches to Causality
        The study of causality in financial time series has long been dominated by linear econometric models, particularly the Granger causality framework and Vector Autoregressive (VAR) models. Granger causality, introduced by \cite{Granger1969}, defines a variable $X$ as a cause of another variable $Y$ if the inclusion of past values of $X$ improves the prediction of $Y$, beyond what is possible using only past values of $Y$. This notion is implemented through VAR models, which represent each variable in a multivariate system as a linear function of its own lags and the lags of other variables in the system \cite{sims1980macroeconomics}.
        However, the reliability of Granger causality and VAR inference critically depends on the assumption of linearity, which is a clear oversimplification in financial time series.
        %several key assumptions, including linearity, stationarity, and homoscedasticity of residuals \cite{hamilton1994time}. 
        %In practice, financial time series often violate these assumptions: relationships among variables may be highly non-linear, influenced by latent factors, or subject to regime changes and structural breaks, see e.g.  \cite{hamilton1989new,bai2003computation,tsay2005analysis}. 
        In this scenario, linear models can produce misleading conclusions about causal relationships or fail to detect them altogether \cite{barnett2009granger,diks2006new}. 
        Recent research has increasingly turned to non-linear and non-parametric methods to analyse causality in financial systems. 
        %These approaches are better suited to capturing the intricate, evolving relationships found in real-world markets, allowing for more nuanced insights into causal dynamics without the restrictive assumptions of traditional linear models. 
        Among these, Transfer Entropy (TE), introduced by \cite{Schreiber200}, stands out as a powerful and model-free measure that detects directional and non-linear causal influences between time series. Unlike Granger causality, which is based on linear predictability, TE quantifies the amount of information flow from the past of one variable to the future of another using an information-theoretic measure.
        %, beyond what can be explained by the target’s own history. 
        %This feature makes TE especially suited to capture subtle asymmetric dependencies in financial markets. 
        Early applications of \cite{baek2005transfer} revealed patterns of sectoral influence in the US stock market, particularly highlighting the role of the energy sector. More recent work, such as \cite{li2018speculative}, applied TE to the Chinese stock market to construct speculative influence networks, demonstrating that sectors exerting stronger influence were more vulnerable during market crashes. In bond markets, \cite{caserini2022effective} used the TE to reveal the leadership of sovereign bonds over credit default swaps in pricing sovereign credit risk during periods of financial distress.
        Despite its theoretical appeal, computing TE in high-dimensional settings is challenging \cite{deltatto2025molecular}, as it relies on estimating probability densities of the dynamic variables.

        Complementing these approaches, Convergent Cross Mapping (CCM), introduced by \cite{sugihara2012detecting}, offers a model-free, state-space reconstruction method to infer directional causality by assessing whether one time series can reliably reconstruct the states of another. CCM has been applied to financial markets, including the work of \cite{ma2024linear}, who identified non-linear causal links between major stock indices during periods of market turbulence.

%% Information-Theoretic and Model-Independent Causal Discovery
        In the context of financial time series, especially under conditions of non-linearity and high dimensionality, approaches that do not rely on the estimation of probability densities offer a powerful alternative to classical tools. A recent and promising framework in this class is Information Imbalance (II), which quantifies the directional predictive power one variable (or set of variables) has over another, using a non-parametric, rank-based comparison of distances \cite{glielmo2022ranking}. In \cite{DelTatto}, the II was employed to detect asymmetric causal relationships between high-dimensional dynamical systems with non-linear dynamics.
        %that signal potential causal structure, making it particularly suitable for financial markets where traditional linear assumptions may fail.

        The Differentiable Information Imbalance (DII) extends this approach by introducing a gradient-based formulation that allows for automated feature selection and weighting, facilitating the application of causal discovery in large and complex datasets \cite{wild2024automatic, DelTatto}. This differentiable variant enhances scalability while preserving the method's non-parametric and model-agnostic nature.
        Compared to other non-linear methods such as TE, 
        %or feature-importance scores from models like Random Forests, 
        the II and DII frameworks offer unique advantages: they do not assume any generative model, can be computed without any explicit estimate of probability densities, and provide directional insights into variable relationships. These properties make them well-suited for causal inference in complex systems, including financial markets. %characterised by feedback, external shocks, and structural dependencies.

        %This potential is demonstrated in recent work by \cite{salvagnin2024investigating}, where Information Imbalance was applied to identify the non-linear drivers of EU Emissions Allowance (EUA) price dynamics, revealing causality patterns that conventional tools failed to detect. The current study builds directly on that foundation, expanding and refining the use of DII for causal discovery across a broader set of variables in the energy and commodity markets.

%% Applications in Energy and Carbon Markets
        The European Union Emissions Trading System (EU ETS) has attracted considerable attention from researchers applying both traditional and modern methods to analyse causality, and risk transmission across energy and financial markets. Numerous studies have employed tools such as Granger causality, spillover indices, and machine learning to uncover the dynamics linking EUA prices with external shocks, compliance cycles, and interconnected markets like energy and equities.

        For instance, Granger causality and volatility models such as EGARCH have been widely used to characterise the temporal dynamics and risk profiles of EUA prices. \cite{Chevallier20111}, and \cite{Chevallier201199}, demonstrate the relevance of conditional volatility and implied option-based measures in capturing regime shifts around compliance events and post-Kyoto uncertainty. High-frequency volatility modelling further reveals asymmetric responses and long-memory features, underscoring the complexity of EUA dynamics \cite{Villar-Rubio2023500}. Similarly, \cite{Conrad2012316} highlight how institutional events and policy milestones affect the stability and predictability of EUA price behaviour.

        More recent work focuses on the role of external shocks and market spillovers. \cite{mi2024} show that geopolitical crises, such as the Russia–Ukraine conflict and the broader energy crisis, substantially amplify volatility in the carbon market through industrial output, emissions demand, and stock-energy linkages. These findings are supported by \cite{Bredin2011353} and \cite{Aslan2022}, who emphasise the structural co-movement between energy prices and EUA volatility due to rising marginal abatement costs and uncertainty. Moreover, the inclusion of sustainability and transition risk proxies, such as energy indices, reflects the growing recognition of systemic risks in low-carbon financial markets \cite{Qiu2023137106}.

        Against this backdrop, \cite{salvagnin2024investigating} offer a novel methodological advance by applying a non-parametric feature selection approach, based on the II, to uncover the variables that are more related to the EUA price formation.
        %directional relationships in the EU ETS. Their work reveals hidden causal drivers in EUA price formation that linear and volatility-based methods may miss, including dynamic feedback from electricity and stock markets. 
        The present study builds upon and extends their methodology by incorporating Differentiable Information Imbalance (DII) into a broader causal framework, showcasing its applicability to high-dimensional market settings.

        %Together, these applications illustrate how volatility modelling, spillover frameworks, and emerging machine learning techniques are converging toward a more granular understanding of causal dynamics in carbon and energy markets, an evolution that this paper seeks to advance.

%% Summary of Gaps and Motivation for the Present Work
        %Despite extensive research on causality in financial and energy markets, existing methods often fall short in capturing the non-linear and directional dynamics of complex high-dimensional systems like the EU ETS. %Traditional linear models (e.g., Granger causality, VARs) struggle with structural breaks, while newer tools, such as spillover indices, machine learning, and transfer entropy, face challenges in interpretability and robustness.
        %This study addresses these gaps by applying Differentiable Information Imbalance (DII), a recent model-agnostic, information-theoretic method that reliably identifies directional, non-linear causal links in complex environments.
        %The key contribution lies in using DII on a multivariate dataset of financial, energy, and carbon variables, offering new insights into EUA price formation and inter-market dependencies amid uncertainty and transition.
%
% Section: Motivation and Goals -----------------------------------------------
    \subsection{Motivation and goals}
    \label{sec1.2:Motivation}
    \noindent
        Despite extensive research on causality in financial and energy markets, existing methods often fall short in capturing the non-linear and directional dynamics of complex high-dimensional systems like the EU ETS.
        This study aims to improve the understanding of energy and commodity market dynamics by proposing a novel approach to identify non-linear causal relationships in financial time series data.

        The approach leverages a non-parametric and non-linear metric called II \cite{glielmo2022ranking}, in its differentiable version DII \cite{wild2024automatic, allione2025linearscalingcausaldiscovery}.
        The II metric used in this work provides a general and robust quantification of the predictive power that one set of variables has on another \cite{glielmo2022ranking}.
%
        %In this sense, this metric generalises traditional methods to quantify predictive power using linear models such as VARs and should not be confused with the concept of information asymmetry, typically used in economics to refer to a situation where the buyer and seller in a contract have different information at their disposal.
%
        %The DII provides a model-independent alternative to established methods, such as Granger causality tests.
%
        To evaluate the effectiveness of the DII, we apply it in conjunction with standard linear methods on a dataset of financial returns, which includes a heterogeneous set of assets and indices related to energy markets.
%
        %The variables are organised into six categories: uncertainty measures, commodities, exchange rates, energy, and country-specific indices.

        A specific objective of this research is to investigate the causal relationships influencing the European Union Allowances market and to examine how external factors affect the financial return dynamics of EUA.
        The impact of each variable is determined by the improvement it brings to the prediction of EUA, compared to a prediction where the same variable is excluded.
        This is quantified using the F-statistic in multivariate Granger causality, and the Imbalance Gain (IG) in the DII approach.

        In our analysis, we assume a condition known as causal sufficiency \cite{runge2018causal_network_reconstruction}, namely that the employed data do not exclude any variable, financial or not, that is a common driver (or confounder) of EUA and the other predictors in the dataset.
        This is an important assumption for both the DII approach and the standard Granger causality analysis.
%
% Although the assumption may not be fully satisfied in practice, we can imagine it to hold at least approximately, given the large and heterogeneous set of financial variables considered in our analysis.
%
        The assumption is most likely not satisfied in practice as unobservable variables such as risk aversion, liquidity needs, expectations about future emissions, economic activity, environmental policies, etc., which contribute to determining prices and returns, are by definition (being unobservable) not included in the dataset.
        Although the estimated causal relationships may not fully account for all confounding influences, they can still reveal dominant interactions and offer interpretable hypotheses for further investigation.

        The current work is closely related to \cite{salvagnin2024investigating}, where the authors introduced the II to identify key determinants of EUA prices.
        That initial investigation inspired the current study, in which we extend the II methodology for causal discovery on EUA returns.
        The key contribution lies in using DII on a multivariate dataset of financial, energy, and carbon variables, offering new insights into EUA price formation and inter-market dependencies amid uncertainty and transition.
%
% Subsection: Organization of the work -------------------------------
    \subsection{Organisation of the work}
    \label{sec1.5:Organisation}
    \noindent
        The rest of this work is structured as follows. 
        Section \ref{sec2:Data} provides details on the dataset used along with descriptive statistics. 
        Section \ref{sec3:Methodology} offers an introduction to the fundamental theoretical background on linear causal discovery (based on VAR models and the Granger test), and on the non-linear causal discovery method we put forward (based on the DII and the IG). The section also explains how the two approaches are related and how they can be applied to our objectives.
        Section \ref{sec4:EmpiricalAnalysis} presents the empirical results obtained by applying multivariate Granger causality and the DII approach to EUA and the variables considered. 
        The discussion of the findings is presented in Section \ref{sec:discussion}, while the concluding remarks are provided in Section \ref{sec6:Conclusions}.
%
% Section: Data -----------------------------------------------
\section{Data}
\label{sec2:Data}
\noindent
    The daily dataset we consider includes a range of market categories such as environmental markets, commodities, exchange rates, energy indices, and country-specific indices.
    The financial returns dataset spans January 2013 to April 2024, totalling 2902 observations.
    In this work, we collect and use the closing prices and hence do not consider any temporal effect that might arise from differences in recording times. 
    As we focus on the long-term relationships between variables rather than on intraday fluctuations, we believe that the absence of precisely aligned recording times does not significantly affect our results.

    Table \ref{tab:returns} presents the dataset, highlighting variables related to uncertainty indicators and commodity prices, notably the VSTOXX volatility index and futures for ICE Brent oil and LME Copper.

       \begin{table}[htp]
        \centering
        \resizebox{0.9\textwidth}{!}{%
        \begin{tabular}{ccccc}
        \hline
        \textbf{ID} & \textbf{Category} & \textbf{Variables} & \textbf{Abbreviations} & \textbf{Database} \\
        \hline
        0 & T & EUA (ICEENDEX) & EUA & Bloomberg\textsuperscript{\textregistered} \\
        \hline
        1 & UNC & GPR & GPR & GPR website \\
        2 & UNC & VSTOXX (V2X) & VSTOXX & Bloomberg\textsuperscript{\textregistered} \\
        3 & UNC & Unc. EUR/USD (CAFZUUEU) & UncEURUSD & Bloomberg\textsuperscript{\textregistered} \\
        4 & UNC & Unc. EUR/JPY (CAFZUEJP) &  UncEURJPY & Bloomberg\textsuperscript{\textregistered} \\
        5 & UNC & Unc. EUR/GBP (CAFZUEGB) &  UncEURGBP &Bloomberg\textsuperscript{\textregistered} \\
        6 & UNC & Unc. EUR/CHF (CAFZUECH) &  UncEURCHF &Bloomberg\textsuperscript{\textregistered} \\
        \hline
        7 & COM & ICE Dutch TTF Natural Gas (TTF0NXHR) & NatGas& Bloomberg\textsuperscript{\textregistered} \\
        8 & COM & Electricity Prices Spain (OMLPDAHD) & ElecES& Bloomberg\textsuperscript{\textregistered} \\
        9 & COM & Electricity Prices Germany (EXAPBDHD) &ElecDE& Bloomberg\textsuperscript{\textregistered} \\
        10 & COM & Electricity Prices France (PWNXFRAV) & ElecFR &Bloomberg\textsuperscript{\textregistered} \\
        11 & COM & ICE Coal Rotterdam futures (TMA Comdty) & CoalFut& Bloomberg\textsuperscript{\textregistered} \\
        12 & COM & LME Copper futures (LMCADS03 Comdty) &CuFut& Bloomberg\textsuperscript{\textregistered} \\
        13 & COM & ICE Brent oil futures (CO1 Comdty) &Brent& Bloomberg\textsuperscript{\textregistered} \\
        14 & COM & Silver (XAG Comdty) & AgFut& Bloomberg\textsuperscript{\textregistered} \\
        15 & COM & Gold (GCZ3 Comdty) & Gold& Bloomberg\textsuperscript{\textregistered} \\
        \hline
        16 & ER & EUR/USD spot (EUR/USD) & EURUSD& Eikon Refinitiv\textsuperscript{\textregistered} \\
        17 & ER & EUR/JPY spot (EUR/JPY) &EURJPY& Eikon Refinitiv\textsuperscript{\textregistered} \\
        18 & ER & EUR/GBP spot (EUR/GBP) &EURGBP& Eikon Refinitiv\textsuperscript{\textregistered} \\
        19 & ER & EUR/CHF spot (EUR/CHF) &EURCHF& Eikon Refinitiv\textsuperscript{\textregistered} \\
        \hline
        20 & ENR & WilderHill New Energy Global Innovation index (Nex) &WHNewEnergy& Bloomberg\textsuperscript{\textregistered} \\
        21 & ENR & Bloomberg Energy TR index (BCOMENTR) &BbgEnergy& Bloomberg\textsuperscript{\textregistered} \\
        22 & ENR & Solactive CEA Future index (SOLAFCEA) & SolCEA & Bloomberg\textsuperscript{\textregistered} \\
        23 & ENR & EUROSTOXX Electricity index (SXEELC) &ESTXElect& Bloomberg\textsuperscript{\textregistered} \\
        24 & ENR & Solactive ESG Fossil EU 50 index &SEF EU50& Bloomberg\textsuperscript{\textregistered} \\
        25 & ENR & Low Carbon 100 index (LC100) & LC100EU& Bloomberg\textsuperscript{\textregistered} \\
        26 & ENR & MSCI Europe Energy Sector index (MXEU0EN) & MSCIEnrg& Bloomberg\textsuperscript{\textregistered} \\
        27 & ENR & ERIX index & ERIX & Bloomberg\textsuperscript{\textregistered} \\
        \hline
        28 & CTRY & Euronext100 (N100) &Euronext100 &Bloomberg\textsuperscript{\textregistered} \\
        29 & CTRY & IBEX35 &IBEX35& Eikon Refinitiv\textsuperscript{\textregistered} \\
        30 & CTRY & DAX &DAX& Eikon Refinitiv\textsuperscript{\textregistered} \\
        31 & CTRY & CAC &CAC& Eikon Refinitiv\textsuperscript{\textregistered} \\
        32 & CTRY & FTSEmib & FTSEmib& Eikon Refinitiv\textsuperscript{\textregistered} \\
        33 & CTRY & Bund 10y EU (BN10) & Bund10y& Bloomberg\textsuperscript{\textregistered} \\
        34 & CTRY & Bond 3m EU (BN03) &Bond3m& Bloomberg\textsuperscript{\textregistered} \\
        \hline
        \multicolumn{5}{c}{\textbf{T}: Target; \textbf{UNC}: Uncertainty variables; \textbf{COM}: Commodity-related variables;} \\
        \multicolumn{5}{c}{\textbf{ER}: Exchange rates; \textbf{ENR}: Energy-related indexes/variables; \textbf{CTRY}: Country indexes.} \\
        \hline
        \end{tabular}
        }
        \vspace{0.3cm}
        \caption{\textbf{Variable descriptions and data sources.} The variables used in the analysis with their full names, abbreviations, codes, categories, and data sources.}
        \label{tab:returns}
    \end{table}
    
    Previous research has demonstrated the importance of emissions trading schemes in influencing market dynamics and price formation \cite{Karpf2018103}. 
    Furthermore, uncertainty indices such as the GPR and VSTOXX have shown strong correlations with market volatility and investment decisions \cite{Baker2016}.
    This dataset also includes spot rates for major currency pairs such as EUR/USD, EUR/JPY, EUR/GBP, EUR/CHF, as well as energy indices, such as the WilderHill New Energy Global Innovation Index.
    The dataset is sourced from Bloomberg and Eikon Refinitiv.
    Standardised returns are used in the following analyses since the variances of the 35 original dataset variables differ by up to 3 orders of magnitude.
    This standardisation ensures comparability and numerical stability. 
    %
    %standardisation is also needed to remove any scale-dependent effect for the considered methods that are not scale-invariant.
    Standardisation is also needed to remove any scale dependence for the weights recovered by both linear and non-linear methodologies, that are not scale-invariant.
    The Imbalance Gain measure, as we will comment in Section \ref{ssec:DII}, satisfies scale-invariance and is not affected by this choice. 
    % ensuring comparability and minimizing influence from larger-scale variables.
%
% Subsection: Computation and Analysis of Financial Returns ------------------------------
    \subsection{Computation and analysis of financial returns}
    \label{ssec:financial_returns}
    \noindent
        The financial returns at time $t$, $R_t$, are calculated as

        \begin{equation}
            R_{t} = \frac{P_{t+1} - P_{t}}{P_{t}},
        \end{equation}
        where $P_t$ is the asset's price at time $t$. 

        Financial returns are commonly used as a proxy for volatility because they reflect price fluctuations, with larger returns typically corresponding to higher volatility. 
        This relationship is central to models namely the ARCH model, where past returns help estimate current volatility \cite{Engle1982}.
        The interest rate variables we consider in our dataset also present negative values.
        Although for time series with negative values, financial returns are often substituted with other transformations \cite{TAY2002847,DIEBOLD2006337}, we prefer to use financial returns for all variables for coherence, as done elsewhere \cite{VICEIRA201297,Andreasen2020}. 
        Furthermore, the mentioned time series change signs very rarely, hence the choice of financial returns does not pose substantial sign problems for this study.

        \begin{figure}[ht]
            \centering
            \includegraphics[width=0.9\textwidth]{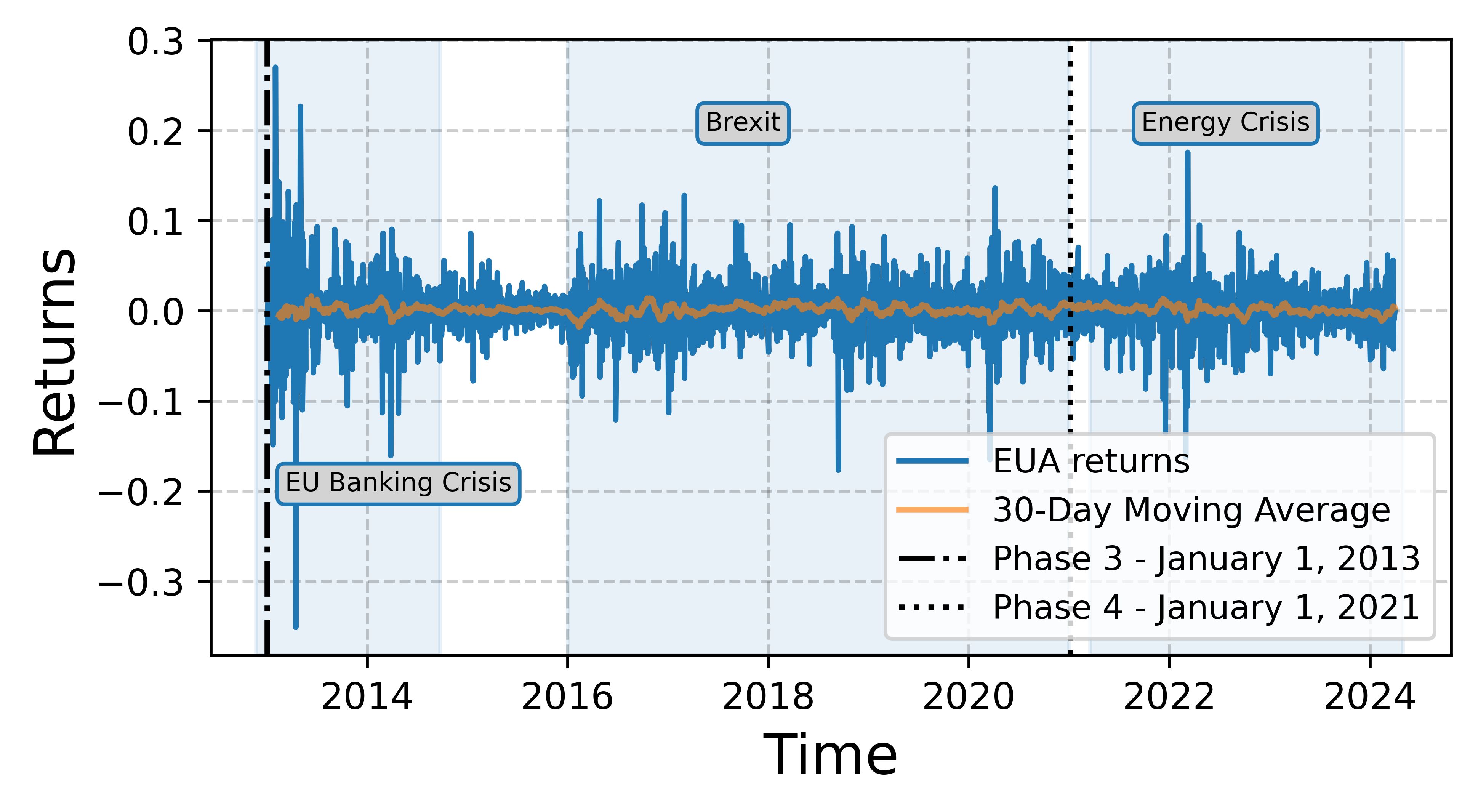}
            \vspace{-0.3cm}
            \caption{\textbf{EUA futures financial returns.} EUA futures returns and some significant events such as the EU Banking Crisis (2012/11 - 2014/09), Brexit (2016/01 - 2020/12), and the Energy Crisis (2021/3 - 2024/4). The returns are smoothed using 30-day moving averages, with key EU ETS phases demarcated. Periods of heightened market uncertainty, reflected by volatility spikes, are aligned with major European crises, which are shaded in light blue for context.}
            \label{fig:rets}
        \end{figure}

        Figure \ref{fig:rets} shows EUA futures returns along with notable economic and geopolitical events that might have affected them. 
        For example, during the EU Banking Crisis (2009–2014), financial instability might have led to reduced liquidity and market confidence, causing firms to reevaluate their emissions and compliance costs. 
        This period was marked by elevated volatility in carbon markets \cite{ecb2022}.

            \begin{table}[htp]
        \centering
        \resizebox{0.9\textwidth}{!}{%
        \begin{tabular}{ccccccccccc}
            \hline
            \textbf{ID} & \textbf{Variables} & \textbf{Mean} & \textbf{STD} & \textbf{Min} & \textbf{Max} & \textbf{25\%} & \textbf{50\%} & \textbf{75\%} & \textbf{Skewness} & \textbf{Kurtosis} \\
            \hline
            0 & EUA  & 0.0013 & 0.0316 & -0.3511 & -0.0143 & 0.0010 & 0.0177 & 0.2703 & -0.2903 & 10.5662 \\
            \hline
            1 & GPR  & 0.0999 & 0.5408 & -0.9500 & -0.2355 & -0.0018 & 0.3031 & 9.4341 & 3.5417 & 37.5951 \\
            2 & VSTOXX  & 0.0024 & 0.0725 & -0.3526 & -0.0398 & -0.0049 & 0.0332 & 0.6256 & 1.3944 & 7.1025 \\
            3 & UncEURUSD  & 0.0001 & 0.0207 & -0.1033 & -0.0078 & 0.0000 & 0.0054 & 0.1850 & 0.9916 & 8.9573 \\
            4 & UncEURJPY  & 0.0002 & 0.0291 & -0.3153 & -0.0076 & 0.0000 & 0.0064 & 0.4380 & 1.6756 & 44.7733 \\
            5 & UncEURGBP  & 0.0003 & 0.0294 & -0.4078 & -0.0081 & 0.0000 & 0.0065 & 0.6793 & 5.7195 & 145.6690 \\
            6 & UncEURCHF  & 0.0011 & 0.0638 & -0.3430 & -0.0065 & 0.0000 & 0.0057 & 3.0505 & 38.3296 & 1814.1401 \\
            \hline
            7 & NatGas  & 0.0009 & 0.0437 & -0.2970 & -0.0155 & -0.0004 & 0.0147 & 0.5110 & 1.4312 & 16.1099 \\
            8 & ElecES  & 0.1398 & 2.1517 & -0.9918 & -0.0755 & -0.0028 & 0.0723 & 70.6610 & 26.3826 & 776.1512 \\
            9 & ElecDE  & 0.0674 & 3.2588 & -72.7895 & -0.1358 & -0.0198 & 0.1348 & 150.2692 & 29.4670 & 1648.0117 \\
            10 & ElecFR  & 0.0467 & 0.7004 & -0.8609 & -0.1047 & -0.0126 & 0.0970 & 32.3447 & 35.1401 & 1574.0379 \\
            11 & CoalFut  & 0.0005 & 0.0259 & -0.3103 & -0.0093 & 0.0001 & 0.0097 & 0.3746 & 0.9355 & 36.1277 \\
            12 & CuFut  & 0.0001 & 0.0120 & -0.0776 & -0.0065 & 0.0000 & 0.0067 & 0.0712 & -0.0431 & 2.2281 \\
            13 & Brent  & 0.0334 & 0.6771 & -0.8609 & -0.0386 & -0.0022 & 0.0249 & 32.3447 & 38.8817 & 1805.5031 \\
            14 & AgFut  & 0.0001 & 0.0176 & -0.1161 & -0.0080 & 0.0000 & 0.0081 & 0.0930 & -0.2531 & 5.5203 \\
            15 & Gold  & 0.0001 & 0.0092 & -0.0907 & -0.0046 & 0.0003 & 0.0050 & 0.0509 & -0.4578 & 6.2462 \\
            \hline
            16 & EURUSD  & -0.0001 & 0.0050 & -0.0278 & -0.0030 & 0.0000 & 0.0026 & 0.0318 & 0.0964 & 2.9530 \\
            17 & EURJPY  & 0.0001 & 0.0060 & -0.0510 & -0.0030 & 0.0001 & 0.0034 & 0.0445 & -0.1842 & 5.4367 \\
            18 & EURGBP  & 0.0001 & 0.0057 & -0.1046 & -0.0026 & -0.0001 & 0.0025 & 0.1164 & 1.2510 & 101.7974 \\
            19 & EURCHF  & -0.0001 & 0.0045 & -0.1689 & -0.0016 & -0.0001 & 0.0016 & 0.0290 & -18.0536 & 688.2632 \\
            \hline
            20 & WHNewEnergy  & 0.0004 & 0.0134 & -0.1023 & -0.0063 & 0.0007 & 0.0071 & 0.0948 & -0.2600 & 5.5705 \\
            21 & BbgEnergy  & -0.0001 & 0.0189 & -0.1353 & -0.0095 & 0.0001 & 0.0097 & 0.1055 & -0.3188 & 4.7386 \\
            22 & SolCEA  & 0.0012 & 0.0314 & -0.3508 & -0.0141 & 0.0005 & 0.0174 & 0.2690 & -0.3441 & 10.8382 \\
            23 & ESTXElect  & 0.0003 & 0.0115 & -0.1594 & -0.0056 & 0.0003 & 0.0066 & 0.0593 & -1.2169 & 15.6282 \\
            24 & SEF EU50  & 0.0002 & 0.0114 & -0.1321 & -0.0049 & 0.0003 & 0.0059 & 0.0881 & -0.7391 & 11.3356 \\
            25 & LC100EU  & 0.0003 & 0.0010 & -0.1100 & -0.0041 & 0.0006 & 0.0052 & 0.0768 & -0.7697 & 9.9159 \\
            26 & MSCIEnrg  & 0.0002 & 0.0164 & -0.1808 & -0.0071 & 0.0004 & 0.0076 & 0.1938 & -0.1182 & 17.7366 \\
            27 & ERIX  & 0.0007 & 0.0163 & -0.1217 & -0.0076 & 0.0010 & 0.0092 & 0.1055 & -0.1282 & 3.8237 \\
            28 & Euronext100  & 0.0003 & 0.0108 & -0.1197 & -0.0045 & 0.0006 & 0.0058 & 0.0818 & -0.7453 & 10.1859 \\
            \hline
            29 & IBEX35  & 0.0002 & 0.0123 & -0.1406 & -0.0060 & 0.0004 & 0.0066 & 0.0857 & -0.9574 & 12.7946 \\
            30 & DAX  & 0.0004 & 0.0118 & -0.1224 & -0.0047 & 0.0006 & 0.0063 & 0.1098 & -0.3781 & 9.6570 \\
            31 & CAC  & 0.0003 & 0.0117 & -0.1228 & -0.0049 & 0.0006 & 0.0061 & 0.0839 & -0.6053 & 9.6292 \\
            32 & FTSEmib  & 0.0003 & 0.0140 & -0.1692 & -0.0062 & 0.0007 & 0.0077 & 0.0892 & -1.1029 & 13.0261 \\
            \hline
            33 & Bund10y  & -0.0080 & 0.7693 & -32.6667 & -0.0348 & 0.0000 & 0.0353 & 13.0000 & -25.8877 & 1165.6719 \\
            34 & Bond3m  & 0.0002 & 0.7113 & -27.0000 & -0.0125 & 0.0000 & 0.0155 & 8.2857 & -18.4459 & 742.9117 \\
            \hline
        \end{tabular}
        }
        \vspace{0.3cm}
        \caption{\textbf{Descriptive statistics of financial returns.} Some key descriptive statistics for the considered financial time series, including the mean, standard deviation (STD), minimum (Min), maximum (Max), skewness, kurtosis, and percentile values (25\%, 50\%, and 75\%).
        }
        \label{Table:DescpritiveStatistics_R}
    \end{table}

        Table \ref{Table:DescpritiveStatistics_R} provides summary statistics for the 35 time series of returns considered. 
        These highlight key features across diverse markets. For instance, EUA futures exhibit a slightly positive mean return (0.0013) with relatively high volatility (0.0316) and negative skewness (-0.2903), indicating a tendency for extreme negative returns. 
        Energy commodities, such as natural gas and Brent oil, show pronounced volatility, driven by geopolitical tensions and supply constraints during the ongoing energy crisis. The Spanish electricity market reveals an even higher dispersion (STD 0.3836), reflecting uncertainty in regional energy markets. Financial indices such as EUROSTOXX and DAX exhibit lower volatility.
       
        These differences underscore the diversity in return behaviours across asset classes. High volatility in energy commodities, extreme kurtosis in currency markets, and asymmetric distributions in financial indices highlight the importance of these characteristics for risk assessment and portfolio management. These patterns emphasise the sensitivity of markets to geopolitical shocks, creating both challenges and opportunities for investors \cite{Baumeister}.
        
        Finally, the post-Brexit increase in the volatility of currency pairs, particularly EUR / USD and EUR / GBP, underscores the economic instability resulting from new trade agreements and regulatory changes \cite{Flori}. 
        Such insights can be important for understanding market dynamics and guiding investment strategies.
        In our analysis, following the causal sufficiency hypothesis, we assume that all economic and geopolitical influences on EUA returns are either directly captured by the variables listed in Table \ref{Table:DescpritiveStatistics_R} or indirectly mediated through one or more variables in our dataset.
        
% Section: Methodology ------------------------------------------------------------------------
\section{Methods}
\label{sec3:Methodology}
\noindent
    In this section, we first review multivariate GC \cite{barrett2010multivariate,blinowska2004granger}, a standard linear method for detecting causal variables and the strengths of their effects, based on VAR.
    We then describe a recently introduced alternative method based on the concept of II, which can be seen as a non-linear generalisation of multivariate GC \cite{allione2025linearscalingcausaldiscovery}.
    Finally, we illustrate the differences and similarities of the two approaches by applying them to two different synthetic datasets.
%
% Subsection: Linear causal discovery: VAR models and Granger's F-statistic ----------------------------------------
    \subsection{Linear causal discovery: VAR models and Granger's F-statistic}
    \label{ssec:GC}
    \noindent
        \subsubsection{VAR models}
        VAR models are often used in conjunction with the GC test to estimate causal weights on financial datasets.
        The VAR model is used to capture the interdependencies between the target and predictor variables over time.
        Consider a VAR($p$) model, where $p$ represents the number of lags, and the system of equations is given by
        \begin{equation}
            \mathbf{v}_t = \sum_{i=1}^p \mathbf{A}_i \mathbf{v}_{t-i} + \mathbf{u}_t,
        \end{equation}
        where $\mathbf{v}_t$ is a vector of variables at time $t$, $\mathbf{A}_i$ is the weight matrix for the $i$-th lag, and $\mathbf{u}_t$ is a white Gaussian random vector. 
        The key idea behind such a VAR is that each variable in $\mathbf{v}_t$ is modelled as a linear function of its past values and the past values of all other variables \cite{hamilton1994time}.

        The vector of variables $\mathbf{v}_t$ is typically assumed to contain both a target variable $z_t$ and potential predictors $\mathbf{x}_t^T=\left(x_t^1,\cdots,x_t^D\right)$.
        Hence, without loss of generality, we can write $ \mathbf{v}_t=(z_t, x_t^1,\cdots,x_t^D)^T$.

        The causal weight $w^\alpha$ associated with the predictor $x^\alpha$ at lag $i$ can be simply computed as the absolute value of the (1,$\alpha$)-component of matrix $\mathbf{A}_i$.
        We recall that in our notation the first row of matrix $\mathbf{A}_i$ contains the weights which determine the impact of all predictors at lag $i$ on the target $z$.
        %The causal weights are calculated by summing the weights corresponding to the predictor variables across all lags.
%
        %Let $\boldsymbol{\beta}_{i}$ denote the row of weight of the matrix $\mathbf{A}_i$ determining the impact of all predictors at lag $i$ on the target variable.
%
        %Then, the causal weight $w^\alpha$ for the predictor $x^\alpha$ is computed as
%   
        %\begin{equation}
        %    w^\alpha = \sum_{i=1}^p (\boldsymbol{\beta}_{i})_\alpha.
        %\end{equation}
%
        The optimal lag for the VAR model can be selected using the Akaike Information Criterion (AIC), which minimises the trade-off between model fit and complexity. 
        The lag with the lowest AIC value is chosen as the optimal one.
        \paragraph{Granger's F-statistic}
        For each time lag, the Granger causality test is used to assess whether the past values of a specific predictor $x^\alpha_{t}$ ($\alpha \in \{1,\cdots,D\}$) help predict the target variable $z_t$ \cite{Granger1969}. 
        The null hypothesis of absence of causality is tested by fitting and comparing two models: a first VAR that includes the lagged values of $x^\alpha_{t}$ in the potential predictors $\mathbf{x}_t$, and a second VAR that does not include $x^\alpha_{t}$, while keeping all other predictors untouched.
        If the test rejects the null hypothesis, it suggests a causal relationship between the predictor $x^\alpha_{t}$ and the target variable $z_t$.
      
        The statistical test relies on the construction of an F-statistic that evaluates the joint significance of the weights associated with the predictor variable, within the VAR model that includes this variable.
        Specifically, for each predictor variable $x^\alpha_{t}$, the F-statistic tests whether the weights associated with the lagged values of $x^{\alpha}_{t}$ across all lags are jointly different from zero.
        A high F-statistic value indicates that the past values of the predictor have significant explanatory power for the target variable, suggesting the presence of a causal relationship \cite{Granger1969}.

        The F-statistic is calculated as
        \begin{equation}\label{eq:f-stat}
            F = \frac{(RSS_r - RSS_u)/q}{RSS_u/(N - k)}.
        \end{equation}
        In the above equation, $RSS_r$ denotes the restricted residual sum of squares, which is obtained from the model where causality restrictions are imposed (i.e., some weights are set to zero). 
        The term $RSS_u$ represents the unrestricted residual sum of squares, calculated from the full model, where all weights are estimated without restrictions. 
        The letter $q$ indicates the number of restrictions, which corresponds to the number of weights set to zero in the restricted model. 
        In standard implementations, the restricted VAR is constructed by setting to zero all lagged values of a single putative causal variable, so $q=p$.
        Finally, $N$ denotes the number of observations, and $k$ the number of parameters estimated in the unrestricted model.
%
        %The numerator $(RSS_r - RSS_u)/q$ represents the reduction in the residual sum of squares due to the restriction i.e., how much better the unrestricted model fits compared to the restricted model, normalised by the number of restrictions.
%
        %The denominator $RSS_u/(N - k)$ is the mean squared error of the unrestricted model, which is an estimate of the variance of the error terms.

        Beyond determining whether to accept or reject the null hypothesis at a chosen significance level, the F-statistic in Eq.~(\ref{eq:f-stat}) can naturally be used to compare the causal impacts of different variables on the same target.
%
% Subsection: The Differentiable Information Imbalance ----------------------------------------
    \subsection{Non-linear causal discovery:  Differentiable Information Imbalance and Imbalance Gain}
    \label{ssec:DII}
    \noindent
    %\paragraph{Information Imbalance}
    \subsubsection{Information Imbalance}
        The methodology employed in this work is a generalisation of the II measure, introduced in~\cite{glielmo2022ranking} and reviewed in the following.
        Given a dataset of $N$ points, the II allows to quantify how much a set of variables $\mathbf{x}$ can predict a given target variable $z$.
        We here assume $\mathbf{x}$ to be multi-dimensional and $z$ to be one-dimensional as this setting is the most relevant for our application, but it is important to stress that the method can be applied regardless of the dimensionality of the two sets.
        The II measure is based on the idea that $\mathbf{x}$ is predictive with respect to $z$ when data points that are close in $\mathbf{x}$ remain close in $z$.
        In practice, we use Euclidean distance functions $d^{\mathbf{x}}$ and $d^z$, such that the distances between two points $i$ and $j$ can be written as $d_{ij}^{\mathbf{x}} = \|\mathbf{x}_i - \mathbf{x}_j \|$ and as $d_{ij}^z = |z_i - z_j |$, where $\mathbf{x}_i$ ($z_i$) denotes the representation of point $i$ in terms of space $\mathbf{x}$ ($z$).
        Then, for each point $i$, we sort distances between $i$ and all the other points $j (\neq i)$ from the smallest to the largest.
        We define the distance rank $r_{ij}^{\mathbf{x}}$ ($r_{ij}^z$) as the position of $d_{ij}^{\mathbf{x}}$ ($d_{ij}^z$) in the list of sorted distances.
        For example, $r^{\mathbf{x}}_{ij} = 1$ if $j$ is the closest point to $i$ according to the distance $d^{\mathbf{x}}$, and $r^{z}_{ij'} = 5$ if $j'$ is the fifth nearest neighbour of $i$ according to $d^z$.
        We write the II from $\mathbf{x}$ to $z$ as
        \begin{equation}\label{eq:II}
            \Delta(\mathbf{x} \rightarrow z) = \frac{2}{N} \mathbb{E}\left[r^z \mid r^{\mathbf{x}} = 1\right] = \frac{2}{N^2} \sum_{\underset{(i\neq j)}{i,j=1}}^N r_{ij}^z \,\delta_{r_{ij}^{\mathbf{x}},\,1}\,,
        \end{equation}
        where $\delta$ denotes the Kronecker delta function, which restricts the sum to pairs of points satisfying $r_{ij}^{\mathbf{x}} = 1$.

        As Eq.~(\ref{eq:II}) shows, $\Delta(\mathbf{x}\rightarrow z)$ is directly proportional to the average distance rank in space $z$, conditioned over pairs of points that are nearest neighbours in $\mathbf{x}$.
        The prefactor $2/N$ allows an asymptotic normalisation of the II to 1, as a function of $N$, in the case of minimum predictivity, namely when $\mathbf{x}$ carries no information about $z$.
        Indeed, in this case $\mathbb{E}\left[r^z \mid r^{\mathbf{x}} = 1\right] = N/2$, as the conditional distance ranks in space $z$ will be uniformly distributed between 1 and $N-1$.

        In the opposite regime, namely when $\mathbf{x}$ is maximally predictive of $z$, $\mathbb{E}\left[r^z \mid r^{\mathbf{x}} = 1\right] = 1$, as all nearest neighbour points in $\mathbf{x}$ will also be nearest neighbours in $z$.
        In this second limit case, $\Delta(\mathbf{x}\rightarrow z) = 2/N$, which approaches 0 in the limit of large $N$.
        Therefore, $\Delta(\mathbf{x}\rightarrow z) \approx 0$ suggests that variables $\mathbf{x}$ are highly informative in predicting variable $z$, while $\Delta(\mathbf{x}\rightarrow z) \approx 1$ indicates that variables $\mathbf{x}$ provide little to no predictive information about variable $z$.

\paragraph{Differentiable Information Imbalance}

        \begin{figure}[htp]
            \centering
            \includegraphics[width=0.9\textwidth]{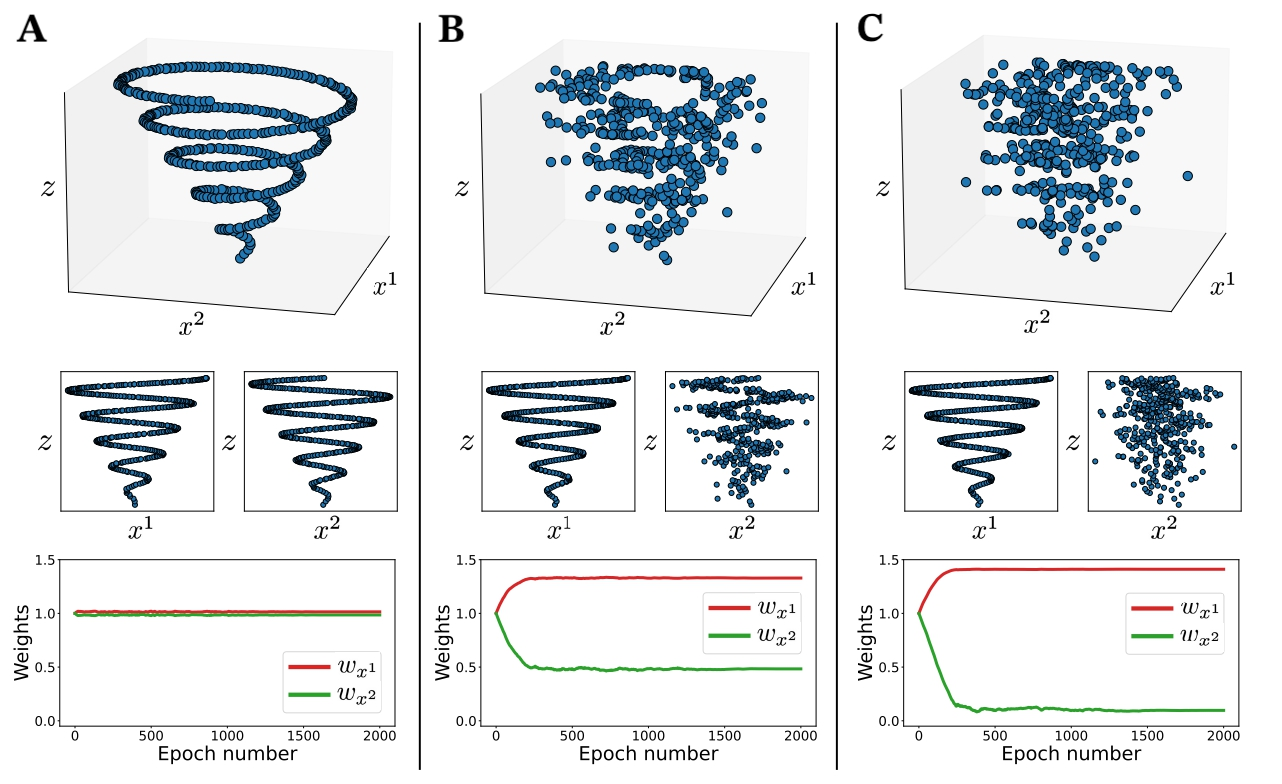}
            \vspace{-0.3cm}
            \caption{\textbf{The Differentiable Information Imbalance.} The DII applied to three spiral datasets with different noise levels. The first row shows the datasets in three-dimensional spaces, while the second row presents projections of the data on two-dimensional planes capturing the marginal dependencies of target variable $z$ on the predictors $x^1$ and $x^2$. The third row shows the value of the weights $w_{x^1}$ and $w_{x^2}$ that the DII assigns to $x^1$ and $x^2$ respectively, as a function of the epoch number in the gradient descent updates. In the left panels, the DII learns that $x^1$ and $x^2$ are equally and jointly important for $z$ as it finds $w_{x^1} \approx w_{x^2} \approx 1$. In central and right panels, the relationship between $z$ and $x^2$ is made progressively more noisy, resulting in the DII weight $w_{x^2}$ shifting towards zero.}
        \label{fig:DII_illustration}
        \end{figure}

        Given a target distance $d^z$, the ability of a set of variables $\mathbf{x}$ to reproduce the neighbourhood relationships defined by $d^z$ depends on the functional form of $d^{\mathbf{x}}$.
        Although the use of the Euclidean distance appears reasonable, as for sufficiently small distances the data manifold can be approximated as locally flat, the relative scale of the different variables entering this distance is to some extent arbitrary.
        This problem was accounted for in \cite{wild2024automatic} by formulating a differentiable version of the II, the DII.
        To introduce it, we define the weighted space $\mathbf{x}(\boldsymbol{w}) = \boldsymbol{w} \odot \mathbf{x} $  with $\odot$ denoting the element-wise product between a vector of arbitrary weights and the space of predictor variables. 
        Each variable in $\mathbf{x}(\boldsymbol{w})$ is scaled by one component of the weight vector $\boldsymbol{w}$.
        Finally, we define the distance $d^{\mathbf{x}}_{ij}(\boldsymbol{w}) = \| \boldsymbol{w} \odot (\mathbf{x}_i - \mathbf{x}_j)\|$ as the Euclidean distance in this weighted space.
        The DII from $\mathbf{x}(\boldsymbol{w})$ to $z$ reads
        \begin{equation}
                \text{DII} \left( \mathbf{x} (\boldsymbol{w}) \rightarrow z \right) = \frac{2}{N^2} \sum_{\underset{(i\neq j)}{i,j = 1}}^N c_{ij}(\lambda, \boldsymbol{w})\, r_{ij}^z\,, \label{eq:DII}
        \end{equation}
        where the weights $c_{ij}$ are defined as
        \begin{equation}
            c_{ij}(\lambda, \boldsymbol{w}) = \frac{e^{-d^{\mathbf{x}}_{ij}(\boldsymbol{w})^2/\lambda}}{\sum_{m(\neq i)} e^{-d^{\mathbf{x}}_{\,im}(\boldsymbol{w})^2/\lambda}}\,. \label{eq:c_ij}
        \end{equation}
        Here, $\lambda$ is a small and positive parameter that defines the size of neighbourhoods in space $d^{\mathbf{x}}(\boldsymbol{w})$.
        Specifically, in the limit $\lambda\rightarrow 0$ the softmax weights in Eq.~(\ref{eq:c_ij}) tend to the Kronecker delta of Eq.~(\ref{eq:DII}), recovering the standard II, where only the nearest neighbour is selected.
        %For optimisation purposes, $\lambda$ is kept in the order of the average squared distance from the nearest neighbour point.
%
        The advantage of the formulation in Eq.~(\ref{eq:DII}) is that derivatives of the DII with respect to weights $\boldsymbol{w}$ can be explicitly computed.
        This allows minimising DII$(\mathbf{x}(\boldsymbol{w})\rightarrow z)$ via gradient descent, namely following an update rule of the form
        \begin{equation}\label{eq:gradient_descent}
            \boldsymbol{w}_{t+1} = \boldsymbol{w}_{t} - \eta\, \nabla_{\boldsymbol{w}} \text{DII}(\mathbf{x}(\boldsymbol{w})\rightarrow z)\,,
        \end{equation}
        until reaching convergence.
        Here, $\eta$ denotes a hyperparameter known as the learning rate and $\nabla_{\boldsymbol{w}} \text{DII}$ is the gradient of the DII with respect to the weight vector.
        As a result, the optimisation of the DII via gradient descent automatically identifies, among the candidate $D$ variables, the weighted combination that optimally predicts the target space $z$.
        An illustrative example is depicted in Figure~\ref{fig:DII_illustration}.
        Further details on the DII optimisation are reported in Appendix \ref{sec:details_DII}.

    \paragraph{Imbalance Gain}
    \label{ssec:IG}
    \noindent
        In \cite{DelTatto} the standard II was employed to detect the presence of causal relationships between high-dimensional dynamical systems, generalising in a non-linear framework the predictability criterion of Granger causality.
    
        Given two one-dimensional time series $x_t$ and $z_t$, one can say that $x_0$ at time $t=0$ improves the prediction of $z$ at a future time $t=\tau$ if a distance measure built with both $x_0$ and $z_0$ is more informative of $z_\tau$ than a distance where only $z_0$ appears.
        This is satisfied when there exists a value of $\tau > 0$ such that the following inequality holds \cite{DelTatto}
        \begin{equation}\label{eq:inequality}
            \min_{w}\Delta\left([w x_0, z_0] \rightarrow z_\tau \right) < \Delta\left( z_0\rightarrow z_\tau \right)\,,
        \end{equation}
        where $[w x_0, z_0]$ is a two dimensional space composed of the scaled variable $w x_0$ and the target variable $z_0$.
        Here, the weight $w$ allows increasing the expressivity of the space containing both $x_0$ and $z_0$, and the minimisation over $w$ is carried out to select the most informative space with respect to $z_\tau$.
        Distances are computed between independent realisations of the dynamics, which can be extracted from a single stationary time series by sampling time frames that are taken as independent initial conditions.
        For processes exhibiting long memory effects, considering time lags $\tau > 1$ can enhance the detection of dynamical couplings, as the impact of a causal link $x \rightarrow z$ may become more pronounced after a specific time span following the interaction \cite{DelTatto, allione2025linearscalingcausaldiscovery}.
    
        The inequality above can be written in terms of a relative difference called Imbalance Gain (IG):
        \begin{equation}\label{eq:IG_std}
            \text{IG}(x\rightarrow z) = 1- \frac{\Delta\left( z_0 \rightarrow z_\tau\right)}{\min_{w}\Delta\left([w x_0, z_0]\rightarrow z_\tau \right)}\,,
        \end{equation}
        which can be expressed as a percentage variation.
        In particular, Eq.~(\ref{eq:inequality}) is equivalent to the condition IG$(x\rightarrow z) > 0$.
        Both equations can be generalised to a multivariate setting, in the same fashion of multivariate GC \cite{DelTatto}.
    
        However, this generalisation comes with additional optimisation parameters, making the minimisation on the right-hand side of Eq.~(\ref{eq:inequality}) computationally demanding.
        In \cite{allione2025linearscalingcausaldiscovery}, this problem was solved using the automatic optimisation of the DII.
        Using the same notation employed in Sec.~\ref{ssec:GC} for GC, the generalisation of Eq.~(\ref{eq:inequality}) in a multivariate setting reads
        \begin{equation}\label{eq:inequality_dii}
            \min_{\boldsymbol{w}}\,\text{DII}\left(\boldsymbol{w}\,\odot\, \mathbf{v}_0\rightarrow z_\tau \right)
            <
            \min_{\boldsymbol{w}'}\,\text{DII}\left( \boldsymbol{w}'\odot (\mathbf{v}_0\setminus x^\alpha_0) \rightarrow z_\tau \right)\,,
        \end{equation}
        where $\mathbf{v}_0 = (z_0, x_0^1, \cdots, x_0^D)^T$, $\boldsymbol{w}$ and $\boldsymbol{w}'$ are vectors of $D+1$ and $D$ optimisation parameters, respectively, and $\mathbf{v}_0 \setminus x_0^\alpha$ denotes vector $\mathbf{v}_0$ without variable $x_0^\alpha$.
        Equivalently, the multivariate version of the IG can be written as
        \begin{equation}\label{eq:IG}
            \text{IG}(x^\alpha \rightarrow z) = 1 -\frac{ \min_{\boldsymbol{w}'}\,\text{DII}\left(\boldsymbol{w}'\odot (\mathbf{v}_0\setminus x^\alpha_0) \rightarrow z_\tau\right)}{\min_{\boldsymbol{w}}\,\text{DII}\left( \boldsymbol{w}\,\odot\, \mathbf{v}_0 \rightarrow z_\tau \right)}\,,
        \end{equation}
        such that a causal effect of $x^\alpha$ on $z$ is detected when $\text{IG}(x^\alpha \rightarrow z) > 0$ for some $\tau >0$.
    
        Importantly, while the optimal weights minimising the DII depend on the variances of the (unscaled) input variables, the minimum value of the DII, and consequently the IG measure, are independent of such original scales.
        Therefore, choosing whether to standardise or not the input variables results in different optimal weights, but does not affect the final estimate of the IG, if equivalent global minima are reached.
%
        %For this reason, we use here the Imbalance Gain as a more reliable measure than the optimal weights to evaluate the actual impact of a predictor variable on the target variable.

    \subsection{Two illustrative applications on synthetic data}
    \label{ssec:synthetic-data}
    \noindent
        We here compare the linear approach described in Sec.~\ref{ssec:GC} with the novel non-linear methodology described in Sec.~\ref{ssec:DII} on two synthetic datasets with a known data-generating process.
        Both datasets were constructed by generating time series
        % of 10$^4$ steps 
        for $x^1,\,x^2$ and $z$, initialising the state of the system with three Gaussian random numbers of unit variance and discarding the initial 5000 steps to avoid equilibration artefacts.
        Then, we selected $N=2800$ time steps after equilibration to match the number of observations used in the empirical analysis of the returns with the DII (see Appendix \ref{sec:details_DII}).
%
%         The analysis with the DII approach was carried out using a dataset of $N=2800$ initial conditions 
%         %AM non capisco cosa voglia dire questa frase sopra, secondo me non è chiara. Forse il verbo SAMPLING non mi è chiaro
%         - the same number employed in the empirical analysis of the returns with the DII approach (see \ref{sec:details_DII}).
% %
%         For the multivariate GC analysis, we considered instead $N=2902$ time steps after equilibration, reflecting the number of observations considered in the returns analysis within the same framework. 
%
        We considered VAR models of order $p=1$ and DII models with time-lag $\tau=1$.
        \subsubsection{False negatives from linear causal discovery}
        The first synthetic dataset is intended to illustrate how the linear causal discovery method can miss important causal predictors that are non-linearly related to a target variable.
        The dataset is constructed by simulating the following stochastic process, with one target variable $z$ and two causal predictors $x^1$ and $x^2$
        \begin{subequations}
        \begin{align}
            x^1_t &= u_t^{1}\,,\\
            x^2_t &= u_t^{2}\,,\\
            z_t &= 0.5 \, z_{t-1} + x^1_{t-1} + (x^{2}_{t-1})^2 + u_t^z. \label{eq:Z}
        \end{align}
        \label{eq:toy-model}
        \end{subequations}
        Here, $u_t^1$, $u_t^2$ and $u_t^z$ are independent Gaussian white noise terms, sampled from $\mathcal{N}(0,1)$.
        According to Eq.~(\ref{eq:Z}), variable $z$ is driven by both $x^1$ and $x^2$, as the state of $z$ at time $t$ is determined by both variables at the previous step.
        Importantly, the causal relationship $x^1\rightarrow z$ is linear, while the link $x^2\rightarrow z$ involves a non-linear (quadratic) coupling.

        \begin{figure}[htp]
            \centering
            \includegraphics[width=0.9\textwidth]{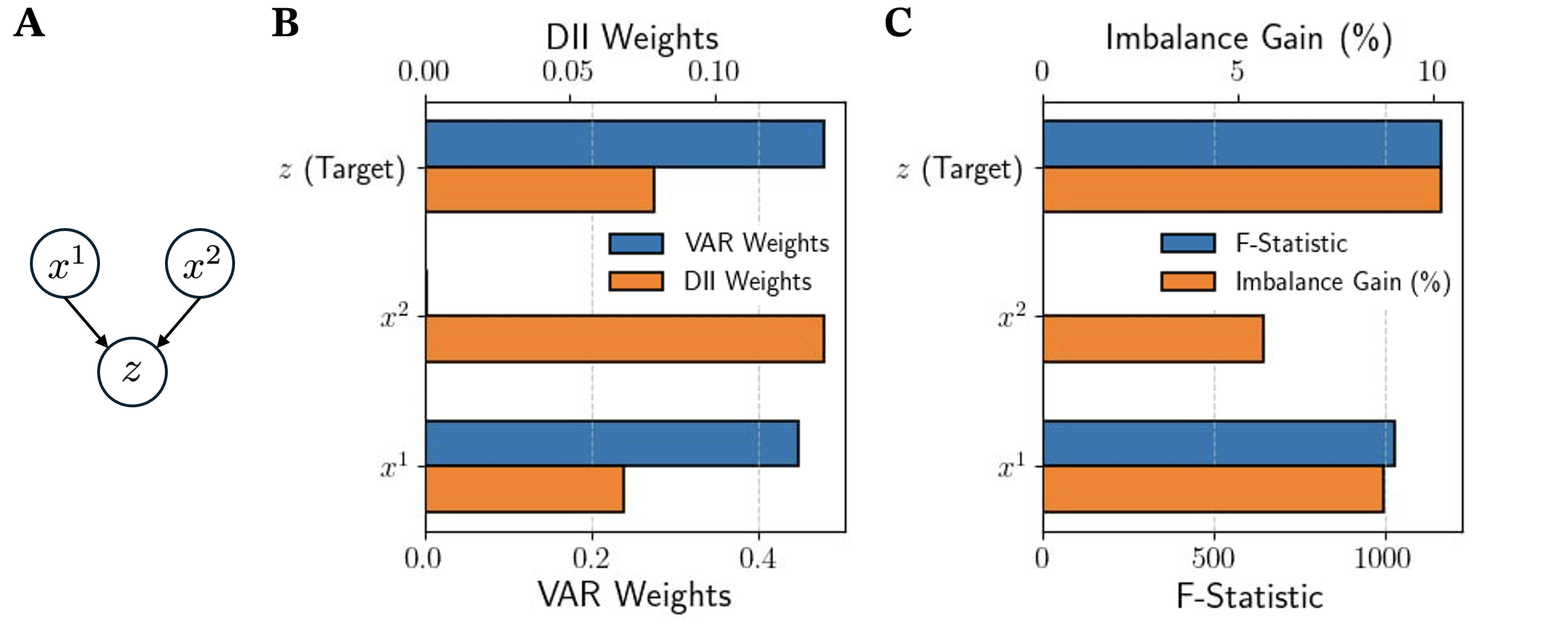}
            \caption{\textbf{False negatives from linear causal discovery.} A comparison of VAR / multivariate GC and DII applied to the stochastic process of Eqs.~(\ref{eq:toy-model}), testing $x^1$ and $x^2$ as putative causal variables with respect to $z$. The graph on the left shows the ground-truth causal relationships among the three variables: $x^1$ and $x^2$ evolve autonomously, while $z$ is caused by both $x^1$ and $x^2$. The bar plot at the centre shows the multiplicative weights assigned to the three variables by fitting the unrestricted VAR(1) model (blue bars) and by optimising the DII$([\mathbf{x}_0,z_0]\rightarrow z_1)$ (orange bars). The right panel shows the F-statistics for each variable, computed within the multivariate GC framework, and the Imbalance Gain obtained from the DII approach.}
            \label{fig:toy-model}
        \end{figure}

        The results of the two methods on this simple stochastic process are reported in Figure \ref{fig:toy-model}.
        As expected, the linear causal relationship $x^1\rightarrow z$ is detected by both methodologies, as can be seen by looking both at the variable weights (left panel) and at the F-statistic and IG measures (right panels).
        On the other hand, the non-linear relationship $x^2\rightarrow z$ is missing according to both the VAR weights and the F-statistic, while it is correctly detected by the DII methodology.
        This simple example shows that standard multivariate GC is prone to false negatives in the presence of non-linear relationships.
%
%%% NEW TOY 2
%
        \subsubsection{False positives from linear causal discovery}
        The second synthetic dataset is intended to illustrate how the linearity assumption of VAR models and multivariate GC can lead to the incorrect detection of false causal relationships.
        The dataset is constructed by simulating the stochastic process
        \begin{subequations}
        \begin{align}
            x_t^1 &= 0.1 \, x^1_{t-1} + (x^{2}_{t-1})^2 + u_t^1, \\
            x^2_t &= 0.7 \, x^2_{t-1} + u_t^2, \\
            z_t &= 0.5 \, z_{t-1} + (x^{2}_{t-1})^2 + u_t^z. \label{eq:Z-2}
        \end{align}
        \label{eq:toy-model-2}
        \end{subequations}
        Here, $u_t^1$, $u_t^2$, and $u_t^z$ are independent Gaussian white noise terms with standard deviations of 0.2, 0.5 and 1.0 respectively.
        In the system described by Eqs.~(\ref{eq:toy-model-2}), $x^2$ is a common driver of $x^1$ and $z$, and causes both variables through a non-linear quadratic relationship.

        \begin{figure}[htp]
            \centering
            \includegraphics[width=0.9\textwidth]{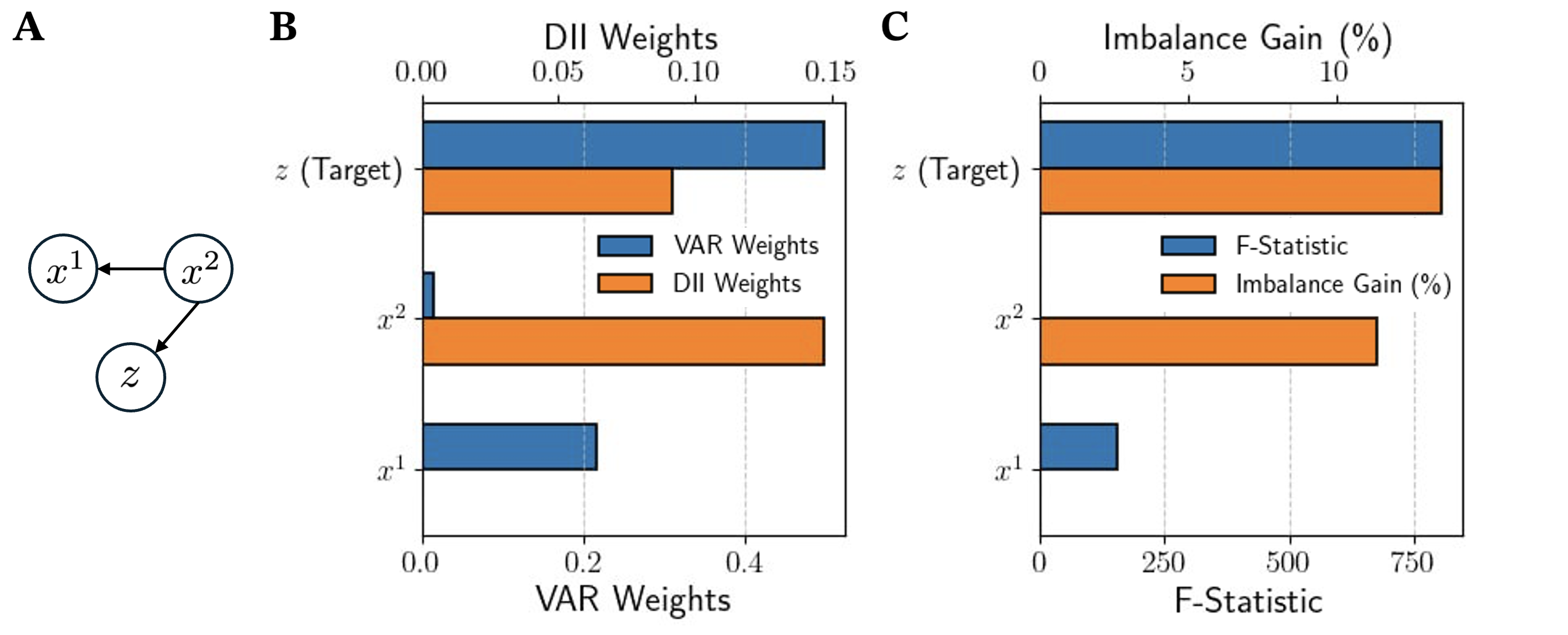}
            \caption{\textbf{False positives from linear causal discovery.} A comparison of VAR / multivariate GC and DII applied to the stochastic process of Eqs.~(\ref{eq:toy-model-2}), testing $x^1$ and $x^2$ as putative causal variables with respect to $z$. The graph on the left shows the ground-truth causal relationships: $x^2$ is a common driver of $x^1$ and $z$.}
            \label{fig:toy-model_2}
        \end{figure}

        Figure \ref{fig:toy-model_2} presents the results of applying multivariate GC (blue bars) and the DII approach (orange bars) to the system under consideration, selecting $z$ as the target variable.
        As in the previous example, we observe that the linear methodology is not able to detect the non-linear coupling $x^2\rightarrow z$, while both the optimal weight from the DII optimisation and the IG measure associated with $x^2$ are significantly non-zero.
        Furthermore, the linear methodology detects a spurious causal link $x^1 \rightarrow z$, as one can see from the magnitudes of both the VAR weight associated to $x^1$ and its F-statistic from the GC test.
        On the other hand, both the DII weights and the IG in direction $x^1\rightarrow z$ allows us to conclude that the impact of $x^1$ on the dynamics of $z$ is actually irrelevant.
%
        %This observation suggests that the Imbalance Gain may be a preferable measure to evaluate the causal impact of each variable.
%
        Our second example illustrates that linear methodologies can suffer from false positive detections in the presence of common drivers and non-linear relationships, while the DII approach appears robust against this drawback.

        The two examples illustrated above motivate the introduction of non-linear generalisations, such as the DII, for analysing real-world time series in which the assumption of linear relationship may be, to different extents, violated.
%
% Section: Empirical Analysis ----------------------------------------------
\section{Empirical analysis}
\label{sec4:EmpiricalAnalysis}
\noindent
    In our analysis of EUA returns, we use the optimal VAR model with lag $p=1$, selected based on the AIC criterion. For the DII analysis, we compare the two methods using a single-day time lag $\tau=1$, consistent with the VAR model. These choices are supported by the ACF and PACF findings (see Tables \ref{Table:Stationarity_Rets}, \ref{Table:VAR_Order_Selection_Rets}, and Figures \ref{fig:acf_rets}, \ref{fig:pacf_rets}). Details regarding DII optimisation are provided in Appendix \ref{sec:details_DII}.
%
    %\subsection{The Granger's F-statistic and the Imbalance Gain}
    %\label{ssec:FStat_IG}
    \noindent

        \begin{figure*}[htp]
            \centering
            \includegraphics[width=0.9\textwidth]{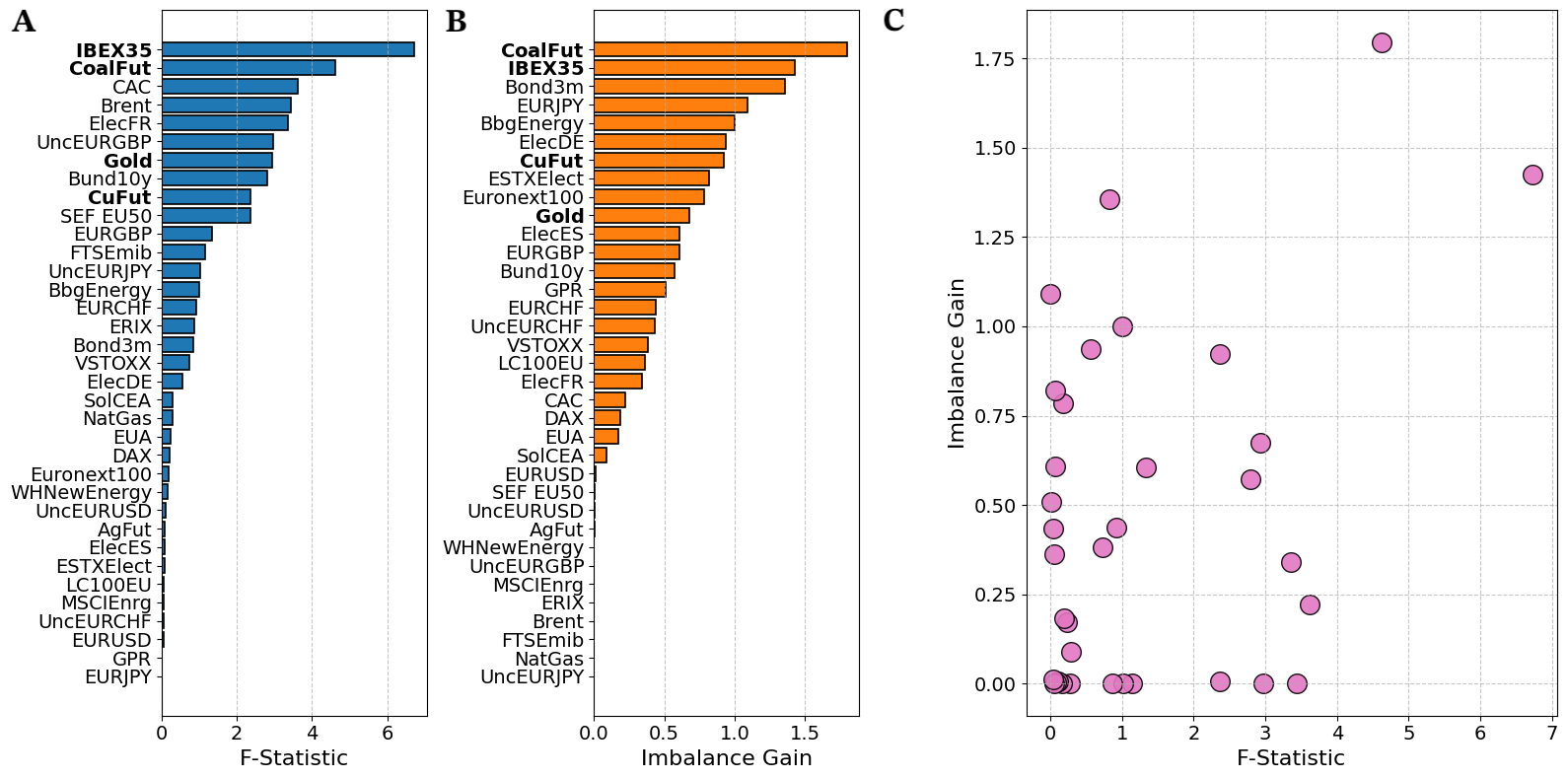}
            \caption{\textbf{The Imbalance Gain as a non-linear alternative to the GC F-statistic.} The results obtained from multivariate GC and the DII approach, using EUA as the target variable. The left panel displays the sorted GC F-statistics associated to each variable, using VAR models of order 1. The central panel displays the sorted IG measures. Bold highlights indicate the top 4 weights among the initial 10 that both methods identify as causal for EUA. The right panel presents a scatter plot correlating the F-statistics and the Imbalance Gain.}
            \label{fig:FStat_IG}
        \end{figure*}

        Figure \ref{fig:FStat_IG} compares the IGs and the Granger's F-statistics, computed for all predictors with respect to the EUA returns.
        In the first panel, the IG highlights IBEX35 index and ICE Coal Rotterdam futures as the variables with the most substantial causal contributions. 
        Interestingly, the IG metric aligns with the Granger's F-statistic in attributing similar importance to these variables. 
        However, a deeper inspection reveals differences in the ranking of causal impacts between the two approaches.
        Among the next eight most important variables, only Gold and LME Copper futures appear as causally relevant for both the linear and the non-linear methods.

        In particular, in the scatter plot of Figure \ref{fig:FStat_IG} we observe that several variables that display a non-zero IG are also assigned a negligible F-statistic.
        Following the findings of the first example of Sec.~\ref{ssec:synthetic-data}, we argue that the causal impact of such variables may be underestimated by multivariate GC as a consequence of non-linear relationships with respect to EUA.

        Indeed, we recall that a key distinction between the two metrics lies in the nature of the couplings they can detect. 
        The F-statistic is sensitive to linear relationships between the target variable and each predictor, which are inherently constrained by the linear structure of the VAR model. 
        In contrast, the IG metric captures non-linear and model-free contributions.
        On the other hand, fewer variables display null IG estimates but non-zero F-statistic.
        
        As illustrated in the second example of Sec.~\ref{ssec:synthetic-data}, this may occur in the presence of a common driver of EUA and the candidate causal variable identified by GC, featuring similar non-linear relationships that are not directly detectable by linear approaches.

        If non-linear relationships play a relevant role in the complex dynamics of financial returns, and in particular in the relationships that affect EUA's evolution, the IG may be a valuable tool to avoid false negatives and false positives produced by standard non-linear techniques.
        While the IG and the F-statistic are similar in their goal of identifying causal relationships, a key difference is that the F-statistic provides a measure of statistical significance, allowing for hypothesis testing regarding the presence of causality. 
        In contrast, IG does not directly offer such a test of significance, as it is based on an unsupervised approach. This difference stems from the fact that the IG methodology does not rely on assumptions of statistical models, making it more flexible but also less suited for significance testing compared to the F-statistic. 
        However, IG provides additional insights by capturing non-linear relationships, which may not be detected using traditional methods like VAR and the F-statistic. To validate the robustness of our findings, we apply both IG and the F-statistic together, ensuring consistency and confirming that IG contributes valuable complementary information.
%
    %\subsection{The VAR weights and the Information Imbalance weights}
    %\label{ssec_VAR_DII}
    %\noindent
%

        \begin{figure*}[ht]
            \centering
            \includegraphics[width=0.9\linewidth]{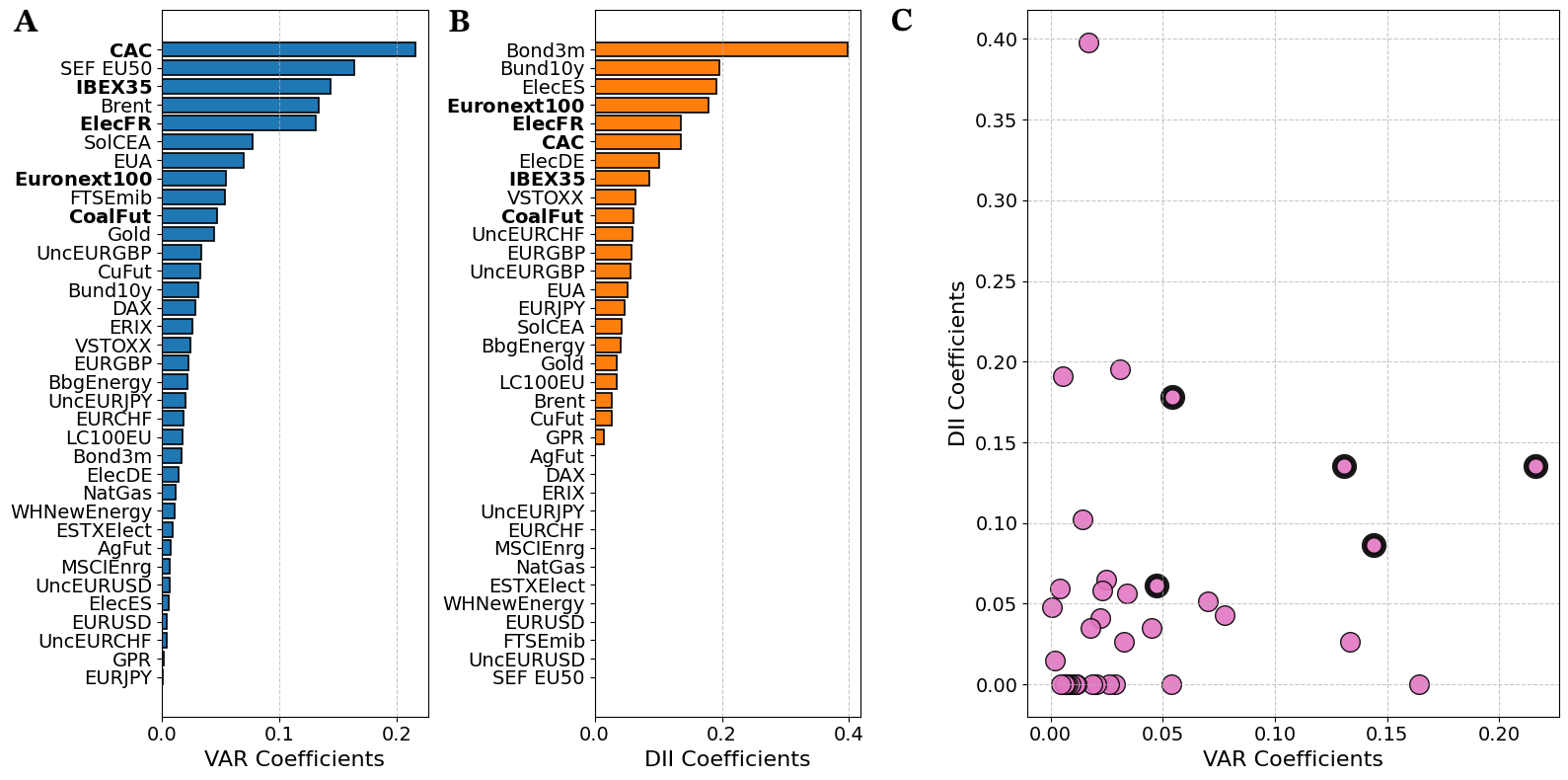}
            \caption{\textbf{Beyond VAR causal weight estimation.} The first panel ranks the estimated weights from a VAR(1) model in decreasing order. The second panel displays causal weights, organised by the DII measure in descending order. Bold highlights indicate the top 5 weights among the initial 10 that both methods identify as causal for EUA. The third panel presents a scatter plot correlating VAR weights with those from the DII measure.}
            \label{fig:VAR_DII}
        \end{figure*}

        To complement the analysis, we report in Figure \ref{fig:VAR_DII} a comparison of weight estimates derived from the VAR(1) model and the DII model.
        Variables highlighted in bold represent the top five weights (out of the ten largest) that are common to the two estimation approaches, showcasing a degree of consistency between the two methods also according to the weight estimates.
        As shown in the applications to synthetic data (Sec. \ref{ssec:synthetic-data}), the optimal weights can provide valuable insights but their estimate in noisy time series could be slightly biased towards non-zero values even in absence of causal links, requiring the use of an appropriate threshold to avoid false positive detections \cite{allione2025linearscalingcausaldiscovery}.
        %do not necessarily measure, in general, the actual impact of the putative causal variables on the target, as non-zero weights may be practically irrelevant due to large uncertainties.
%
        %Therefore, we regard the results obtained from the IG as more reliable and 
        In this work, we present the optimal DII weights solely for comparison with their linear counterparts.

% Section: Discussion ------------------------------------------------
\section{Discussion on the variables identified}
\label{sec:discussion}
% High level introduction -----------------------------------------------------
    The EU ETS market operates within a highly interconnected network of European markets, where various economic and energy-related drivers influence the prices of the European carbon Allowances (EUA).
    
    Our study employs the novel DII measure, along with a Granger VAR(1) model, to identify causal relationships. As demonstrated in Sec.~\ref{sec4:EmpiricalAnalysis}, both approaches consistently highlight two key drivers: the influence of the IBEX35 Spanish stock index and European coal futures on EUA financial returns.

% Economic comment on IBEX35 -------------------------------------------------
    The evidence showed a Granger-causal link between the Spanish stock market and EUA financial returns that indicates that fluctuations in IBEX35 index can predict changes in carbon allowance prices.
    Following \cite{JIMENEZRODRIGUEZ201913}, this relationship was especially pronounced during Phase II of the EU ETS, which coincided with the Global Financial Crisis (2007-2008). During this period, the financial crisis led to reduced economic activity, lower industrial production, and a contraction in energy demand. As a result, firms were more likely to sell their EUAs to increase liquidity, which in turn had significant effects on carbon prices \cite{BEL2015531}.
    In general, the causal relationship between stock market movements and EUA price changes can be explained by broader economic shocks, where equity markets often reflect investor sentiment regarding future economic conditions, which in turn affects demand for carbon allowances \cite{calel2012environmental}.
    
% Economic comment on Coal Futures -------------------------------------------
    Consistently with our findings, \cite{su9101789} identified strong volatility links between coal futures and EUA, highlighting the close interrelationship between commodities and EU ETS.
    These results suggest that fluctuations in coal prices, as a major energy source with significant carbon emissions, exert a direct influence on carbon allowance prices.
    The interconnectedness arises from the fact that coal is often a key input in industries that must comply with carbon emission regulations, suggesting that changes in coal prices can affect the cost of carbon allowances \cite{Ji2018972}.
    Consequently, shifts in energy prices, particularly those driven by the cost of fossil fuels such as coal, tend to ripple through carbon markets, influencing EUA pricing dynamics. This mutual influence on prices reflects the broader economic mechanisms at play, where energy market shocks often drive changes in emissions-related costs and vice versa \cite{ZHANG20162654}.

% Section: Conclusions -----------------------------------------------
\section{Conclusions}
\label{sec6:Conclusions}
\noindent
%%%%%% HIGH LEVEL SUMMARY %%%%%%
    In this work, we introduced a novel approach for identifying non-linear causal relationships within energy and commodity markets, specifically focussing on the European Union Allowances market.
    The method we proposed is based on the Information Imbalance, a non-parametric and model-free measure to quantify the non-linear predictive power that one space possesses on another.
    More specifically, we leveraged its differentiable version, the Differentiable Information Imbalance, as an alternative to traditional linear models such as VARs for detecting causal relationships.
    As in VAR models, the DII approach assigns a single weight to each variable in the dataset, allowing for a one-to-one comparison of the results from the two methodologies.
    In addition, we proposed the Imbalance Gain measure as an alternative to the Granger's F-statistic for evaluating causal contributions.
    Based on the DII model, the Imbalance Gain can capture both linear and non-linear relationships in a model-free context.
    
%%%%%% TOY-MODELS SUMMARY %%%%%%
    Using simple stochastic processes with known underlying equations, we demonstrated that the DII approach can accurately detect the underlying couplings, even when VAR and multivariate GC fail due to the presence of non-linear relationships.
    The Imbalance Gain measure, in particular, proved valuable in identifying non-linear causal links that the F-statistic fails to capture, potentially offering a more precise representation of the variable impacts.
    Furthermore, we showed that linear methodologies may result in spurious causal detections due to the presence of common drivers and non-linear relationships, while the Imbalance Gain can reliably distinguish real from spurious links in the same scenario.

%%%%%% REAL-DATA SUMMARY %%%%%%
    We then analysed a heterogeneous dataset of financial returns from January 2013 to April 2024. 
    Here, our empirical analysis revealed that the DII approach and the VAR approach show some overlap in identifying key variables influencing EUA returns. 
    Specifically, both methodologies detect significant causal effects on the EUA returns from IBEX35 and Coal Futures.
    %
    %These findings align with those of \cite{JIMENEZRODRIGUEZ201913} and \cite{su9101789}.
    %
    %Evidence of unidirectional Granger causality from the IBEX35 to EUA prices suggests that movements in the Spanish stock market (IBEX35) predict carbon allowance price changes, particularly during Phase II, which coincided with the financial crisis. 
    %
    %This period saw intensified impacts due to reduced economic activity, lower energy demand, and companies selling EUAs for liquidity.
%
    %Similarly, \cite{su9101789} identified strong volatility linkages between coal and EUA futures, indicating that energy and carbon markets are tightly interconnected and with mutual price influences.

%%%%%% HIGH-LEVEL IMPORTANCE %%%%%%
    By introducing the Imbalance Gain as an analogous alternative to the F-statistic, and the DII model as an analogous alternative to the VAR model, we have extended the tools available for causal discovery, with a non-linear and model-free technique.
    This is beneficial for two reasons.
    First, the DII methodology can be used to validate results coming from standard VAR-based approaches.
    Indeed, the agreement of DII-based and VAR-based causality detection on a few variables, such as IBEX35 and Coal Futures in our study, can be taken as strong evidence that such variables are causally relevant.
    Second, as demonstrated on synthetic data, the DII methodology is more robust and effective than standard linear methods when significant non-linear relationships come into play \cite{cont2001, hsieh1991}.

%%%%%% FUTURE WORK %%%%%%
    In this respect, we foresee different complementary lines of future investigation.
    To start with, it would be important to further investigate the empirical discrepancies found between the results of VAR and DII methodologies.
    While we demonstrated on two synthetic datasets that the differences might be due to the intrinsic limitations of the VAR model, these differences require further exploration to better understand how linear methods, like VAR, might miss or misrepresent certain causal relationships that could be better captured by non-parametric methods such as DII.
    On the other hand, systematic studies on the robustness of the DII approach against different levels of noise and sampling times may be beneficial for assessing its reliability in challenging real-world scenarios. 
    An important limitation of the DII approach is the lack of a simple way to estimate the statistical significance or the confidence intervals for its prediction.
    This is not a limitation of the linear methods considered, since VAR weights and F-statistics can easily be assigned to a given statistical significance.
    Future research should focus in this direction to develop an efficient methodology to compute the statistical significance of the predictions coming from DII approach using, e.g., efficiently implemented resampling techniques.
    %
    % Commento MT: "I think this discussion could be more nuanced. With the VAR you can do statistical inference (assess how statistically significant causality is), which you cannot do with your model (or you can do it numerically / with bootstrap at a very steep cost)."
    %
    % Furthermore, in \cite{DelTatto} it was empirically observed that the use of time-delay embeddings, namely the selection of more time frames for each predictor, allows to recover meaningful results even in the presence of unobserved variables.
    % %
    % Within the DII framework, the extension of this approach may be valuable to relax the hypothesis of causal sufficiency, which may be violated in real systems such as a financial market.
%
    Another line of future research could involve leveraging the DII model to extend the Spillover methodology \cite{Diebold2012} with a non-linear analogue, and comparing the causal networks derived from the VAR Spillover, with those generated by the DII approach \cite{allione2025linearscalingcausaldiscovery}.
    While the VAR Spillover model offers a solid framework for analysing linear relationships and dynamic interactions, the DII approach could provide a more flexible tool that can better adapt to non-linear dependencies.
    This comparison will help clarify the advantages of non-parametric causal inference methods and could provide a more robust framework for understanding causal interactions in financial and economic data.
    
    Finally, although we applied the DII methodology to financial data related to EUA market, the technique we proposed is general and could be applied seamlessly to other financial markets.
    This could provide further evidence of the value of the DII methodology, proving it as a valuable new tool for robust causal discovery.
\newpage
%
% Section: Acknowledgments --------------------------------------------------------------------
\section*{Acknowledgments}
\label{sec:Acknowledgments}
   \noindent
    We sincerely thank Matteo Allione (Politecnico di Torino, Italy) for his valuable collaboration on the project. His support and contributions have been greatly appreciated throughout the development and application of the methodology.
    We thank Luigi Bellomarini, Marco Benedetti, Claudia Biancotti, Ivan Faiella and Marco Taboga (Banca d'Italia, Italy) for their useful feedback on this work.
%
% Section: Funding ----------------------------------------------------------------------------
\section*{Funding}
\label{sec:Funding}
   \noindent
    The work by M. E. De Giuli has been supported by the Italian Minister of University and Research (MUR) project: A geo-localized data framework for managing climate risks and designing policies to support sustainable investments (No. 20229CWYXC) within the PRIN 2022 program. Her research was also supported by the Centre for the Analysis and Measurement of Global Risks (CAM-Risk) Project Financial Oversight and Risk-Tailored Understanding for New Evaluation.
    \\
    \noindent
    The work by A. Mira has been supported by Swiss National Science Foundation Grant 200021\_208249.
%
% Section: CRediT authorship contribution statement -------------------------------------------
\section*{CRediT authorship contribution statement}
\label{AuthorContributions}
    \noindent
    \textbf{Cristiano Salvagnin:} Conceptualisation, Software, Validation, Formal analysis, Investigation, Data Curation, Visualisation, Writing - Original draft, Writing - Review and Editing. \textbf{Vittorio Del Tatto:} Methodology, Software, Formal analysis, Validation, Visualisation, Writing - Review and Editing. \textbf{Maria Elena De Giuli:} Writing - Review and Editing, Supervision, Funding acquisition. \textbf{Antonietta Mira:} Conceptualisation, Writing - Review and Editing, Supervision, Funding acquisition. \textbf{Aldo Glielmo:} Conceptualisation, Methodology, Software, Writing - Review and Editing,  Supervision, Project Administration.

% Section: Disclaimer -------------------------------------------------------------------------
\section*{Disclaimer}
\label{sec:Disclaimer}
\noindent
    The views and opinions expressed in this paper are those of the authors and do not necessarily reflect the official policy or position of Banca d’Italia.
%
% Section: Disclosure statement -----------------------------------
\section*{Disclosure statement}
\label{DeclarationOfInterest}
    \noindent
    The authors report there are no competing interests to declare.
%
% Section: Replicating and supplementary materials --------------------------------------------
\section*{Replicating and Supplementary Materials}
\label{sec:ReplicatingSupplementaryMaterials}

    \noindent
    All replication materials, supplementary information, and datasets are available in a dedicated GitHub repository titled \textit{EUA Causal Discovery DII}, accessible at:\\
    \texttt{https://github.com/SaveChris/EUACausalDiscoveryDII}

    \vspace{0.5em}
    \noindent
    The implementation of Differentiable Information Imbalance is based on the DADApy software package \cite{glielmo2022dadapy}.
%
% Bibliography -----------------------------------
\bibliographystyle{plain}  % Neutral bibliography style
\bibliography{arXiv_refs}  % External .bib file

% Appendix --------------------------------------------------
%
\appendix

% A.1 Section:Details on the DII optimisation --------------
    \section{Details on the DII optimisation}
    \label{sec:details_DII}
    \noindent
    We report in this section technical details about the DII optimisation.
    A relevant hyperparameter of the method is $\lambda$ appearing in the softmax coefficients of Eq.~(\ref{eq:c_ij}), which defines the size of the neighbourhoods in the first distance space.
    Although the classical II is recovered in the limit $\lambda \rightarrow 0$, too small values of $\lambda$ make the DII optimization inefficient, as decreasing $\lambda$ also decreases the magnitude of the DII derivatives employed in the gradient descent updates \cite{wild2024automatic}.
    In this work, we compute $\lambda$ with a point-adaptive scheme. Namely, we use a distinct value $\lambda_i$ for each set of coefficients $\{c_{ij}\}_{j=1}^N$.
    This approach allows handling data sets where points are not homogenously distributed, and where a single distance scale would result in different numbers of neighbours for points in different regions of the data manifold.
    Specifically, we computed each $\lambda_i$ as
    \begin{equation}
        \lambda_i = 0.1\, d^{\mathbf{x}}_{ij(k)}(\boldsymbol{w})^2\,,
    \end{equation}
    where 0.1 is an empirical prefactor and $j(k)$ is the $k$-th nearest neighbour of $i$ according to distance $d^{\mathbf{x}}(\boldsymbol{w})$.
    In all the optimisations carried out in this work, $k$ was fixed to 5\% of the points entering the calculation of the DII (i.e., $k/N = 5\%$).
    To keep into account the variation of $d^{\mathbf{x}}(\boldsymbol{w})$ during the DII optimisation, the set of $\{\lambda_i\}$ is recomputed at each gradient descent update.

    Other relevant aspects that may affect the convergence of the DII to its global minimum are the choice of the optimiser and the use of mini-batches.
    The use of mini-batches, namely the computation of the gradient on random subsets of points at each gradient descent update, is a well-known strategy that can improve convergence speed and stability of training in optimisation algorithms, especially when the loss function features several local minima.
    In this work, all the optimisations were carried out by selecting $N=2800$ distinct frames from the original time series, and randomly splitting the data set into 28 mini-batches with $N'=100$ points in each training epoch.
    Using this approach, the value of $k$ defining the set of $\{\lambda_i\}$ was fixed to 5 in each mini-batch.
    Beyond improving the convergence properties, in our application the use of small mini-batches allows to sample time frames which are likely to be uncorrelated, satisfying one of the requirements of our approach (see Sec.~\ref{ssec:IG}).
    In contrast, computing the DII without mini-batches would require an undersampling of the original time series to fulfil this requirement, reducing the statistical significance of the results when dealing with short time series.
    Rather than using the vanilla gradient descent approach provided in Eq.~(\ref{eq:gradient_descent}), in this work we used the Adam optimiser \cite{kingma2017adam}, which is widely employed by the machine learning community and well-known for its good convergence properties.

    All optimisations were carried out for 2000 training epochs, setting the initial learning rate $\eta$ to $10^{-3}$, and progressively decreasing it to zero according to a cosine decay schedule.
    After learning the optimal weights, the DII terms appearing in the IG expression of Eq.~(\ref{eq:IG}) were computed over all the $N=2800$ time frames originally extracted.
    In the analysis of the financial returns, in order to satisfy the requirement of independent initial conditions in the DII calculation, the final DII estimates were carried out by discarding, for each frame $t$, distances with points within a time window $\left[t-1,t+1\right]$.
    Similarly, in both systems of Sec.~\ref{ssec:synthetic-data} the final IGs were computed by discarding distances between each frame $t$ and points within $\left[t-5, t+5\right]$.
%
% A.2 Section: Stationarity and Optimal VAR Lag Order ------
    %\clearpage
\section{Stationarity and Optimal VAR Lag Order}
\label{sec:stationarity}
        \begin{table}[ht]
            \centering
            \resizebox{0.9\columnwidth}{!}{%
            \begin{tabular}{lcccccc}
            \hline
            Asset & ADF Statistic & p-value & Stationary & 1\% & 5\% & 10\% \\
            \hline
            EUA & -15.038364 & \num{9.674195e-28} & True & -3.432621 & -2.862543 & -2.567304 \\
            GPR & -39.378217 & \num{0e+00} & True & -3.432621 & -2.862543 & -2.567304 \\
            VSTOXX & -12.997073 & \num{2.735961e-24} & True & -3.432621 & -2.862543 & -2.567304 \\
            UncEURUSD & -9.957579 & \num{2.421169e-17} & True & -3.432621 & -2.862543 & -2.567304 \\
            UncEURJPY & -25.692109 & \num{0e+00} & True & -3.432621 & -2.862543 & -2.567304 \\
            UncEURGBP & -28.913692 & \num{0e+00} & True & -3.432621 & -2.862543 & -2.567304 \\
            UncEURCHF & -45.237371 & \num{0e+00} & True & -3.432621 & -2.862543 & -2.567304 \\
            NatGas & -9.492228 & \num{3.626879e-16} & True & -3.432621 & -2.862543 & -2.567304 \\
            ElecES & -20.773602 & \num{0e+00} & True & -3.432621 & -2.862543 & -2.567304 \\
            ElecDE & -54.043368 & \num{0e+00} & True & -3.432621 & -2.862543 & -2.567304 \\
            ElecFR & -16.706846 & \num{1.445623e-29} & True & -3.432621 & -2.862543 & -2.567304 \\
            CoalFut & -9.772322 & \num{7.082209e-17} & True & -3.432621 & -2.862543 & -2.567304 \\
            CuFut & -19.373083 & \num{0e+00} & True & -3.432621 & -2.862543 & -2.567304 \\
            Brent & -16.711492 & \num{1.433337e-29} & True & -3.432621 & -2.862543 & -2.567304 \\
            AgFut & -10.018897 & \num{1.699656e-17} & True & -3.432621 & -2.862543 & -2.567304 \\
            Gold & -30.521431 & \num{0e+00} & True & -3.432621 & -2.862543 & -2.567304 \\
            EURUSD & -18.132932 & \num{2.503117e-30} & True & -3.432621 & -2.862543 & -2.567304 \\
            EURJPY & -56.465380 & \num{0e+00} & True & -3.432621 & -2.862543 & -2.567304 \\
            EURGBP & -29.500339 & \num{0e+00} & True & -3.432621 & -2.862543 & -2.567304 \\
            EURCHF & -17.840216 & \num{3.108304e-30} & True & -3.432621 & -2.862543 & -2.567304 \\
            WHNewEnergy & -12.735846 & \num{9.164802e-24} & True & -3.432621 & -2.862543 & -2.567304 \\
            BbgEnergy & -55.477960 & \num{0e+00} & True & -3.432621 & -2.862543 & -2.567304 \\
            SolCEA & -15.054414 & \num{9.194347e-28} & True & -3.432621 & -2.862543 & -2.567304 \\
            ESTXElect & -16.622364 & \num{1.693681e-29} & True & -3.432621 & -2.862543 & -2.567304 \\
            SEF EU50 & -13.901380 & \num{5.726032e-26} & True & -3.432621 & -2.862543 & -2.567304 \\
            LC100EU & -11.741741 & \num{1.268883e-21} & True & -3.432621 & -2.862543 & -2.567304 \\
            MSCIEnrg & -15.784380 & \num{1.121322e-28} & True & -3.432621 & -2.862543 & -2.567304 \\
            ERIX & -17.304976 & \num{5.599100e-30} & True & -3.432621 & -2.862543 & -2.567304 \\
            Euronext100 & -11.347695 & \num{1.017465e-20} & True & -3.432621 & -2.862543 & -2.567304 \\
            IBEX35 & -36.352004 & \num{0e+00} & True & -3.432621 & -2.862543 & -2.567304 \\
            DAX & -17.000522 & \num{8.735585e-30} & True & -3.432621 & -2.862543 & -2.567304 \\
            CAC & -11.317838 & \num{1.194384e-20} & True & -3.432621 & -2.862543 & -2.567304 \\
            FTSEmib & -37.865048 & \num{0e+00} & True & -3.432621 & -2.862543 & -2.567304 \\
            Bund10y & -13.497101 & \num{3.026902e-25} & True & -3.432621 & -2.862543 & -2.567304 \\
            Bond3m & -12.556939 & \num{2.144102e-23} & True & -3.432621 & -2.862543 & -2.567304 \\
            \hline
            \end{tabular}
            }
            \vspace{0.3cm}
            \caption{\textbf{Augmented Dickey-Fuller (ADF) Test Results.} The table presents the results of the ADF test applied to the financial returns dataset. The ADF test checks for stationarity by testing the null hypothesis that the series has a unit root. A p-value below 0.05 indicates stationarity, allowing rejection of the null hypothesis.}
            \label{Table:Stationarity_Rets}
        \end{table}

        \begin{table}[ht]
            \centering
            \small
            \begin{tabular}{|c|c|c|c|c|}
            \hline
            \textbf{Lag} & \textbf{AIC} & \textbf{BIC} & \textbf{FPE} & \textbf{HQIC} \\
            \hline
            1 & \textbf{-30.11}* & -27.51* & 8.377e-14* & -29.17* \\
            2 & -29.97 & -24.84 & 9.680e-14 & -28.12 \\
            3 & -29.71 & -22.05 & 1.250e-13 & -26.95 \\
            4 & -29.56 & -19.37 & 1.462e-13 & -25.89 \\
            5 & -29.55 & -16.83 & 1.482e-13 & -24.96 \\
            6 & -29.21 & -13.96 & 2.091e-13 & -23.71 \\
            7 & -28.89 & -11.12 & 2.870e-13 & -22.49 \\
            8 & -28.58 & -8.277 & 3.960e-13 & -21.26 \\
            9 & -28.38 & -5.550 & 4.875e-13 & -20.15 \\
            10 & -28.35 & -2.989 & 5.091e-13 & -19.21 \\
            \hline
            \end{tabular}
            \vspace{0.3cm}
            \caption{\textbf{VAR Model Order Selection using AIC, BIC, FPE, and HQIC.} The table shows the optimal lag selection for the VAR model based on several criteria. The minimum values for each criterion are highlighted with an asterisk. The lag order with the lowest AIC value is considered optimal.}
            \label{Table:VAR_Order_Selection_Rets}
        \end{table}
        
        \begin{figure*}[ht]
        \centering
        \includegraphics[width=0.9\textwidth]{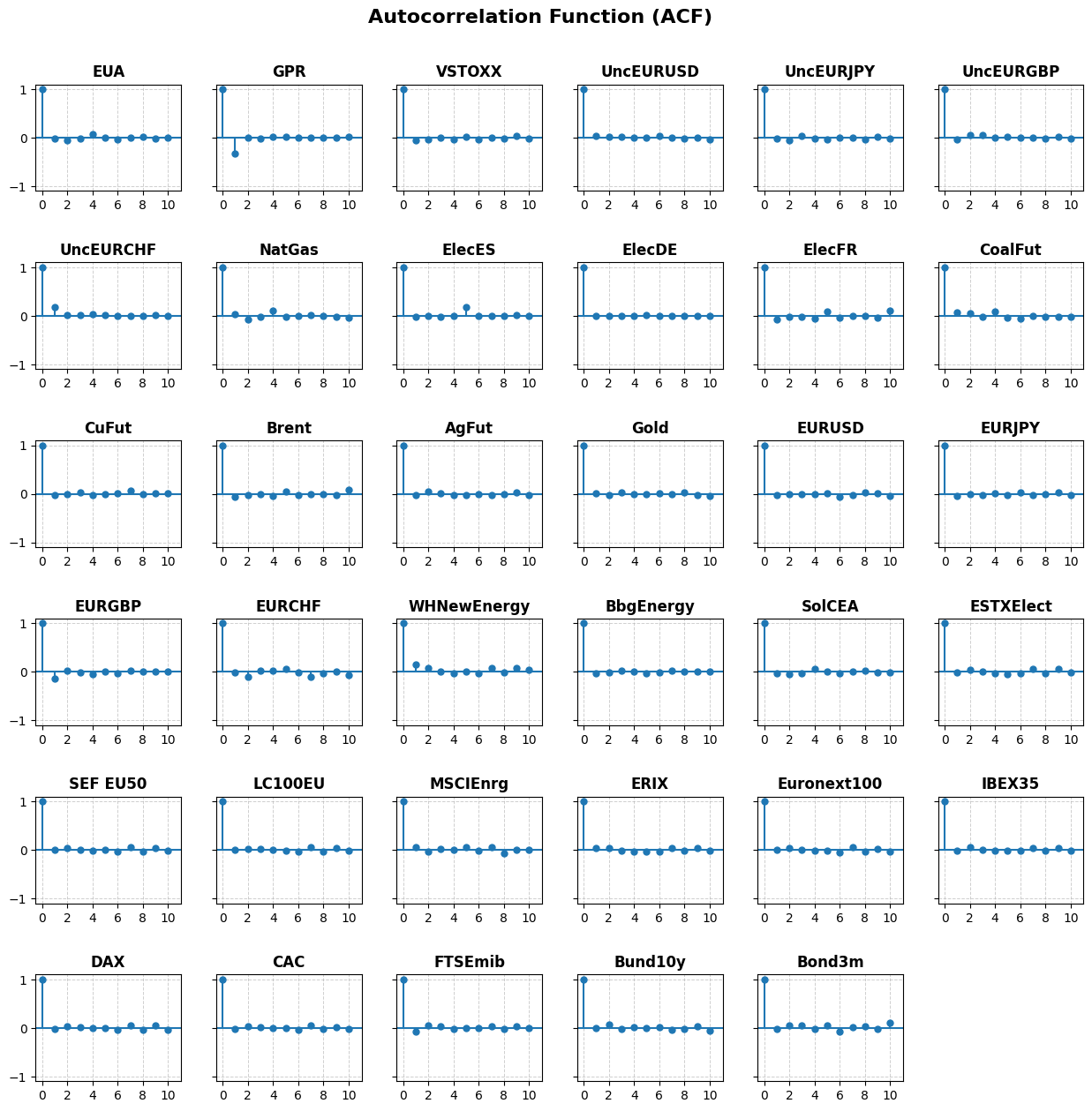}
        \caption{\textbf{Autocorrelation Function (ACF) Plot of Financial Returns.} The ACF plots show very fast decays, exhibiting values very close to zero already from lag 1.
        This aligns with the widely recognized understanding that predicting financial returns is exceedingly difficult.}
        \label{fig:acf_rets}
        \end{figure*}
        
        \begin{figure*}[ht]
        \centering
        \includegraphics[width=0.9\textwidth]{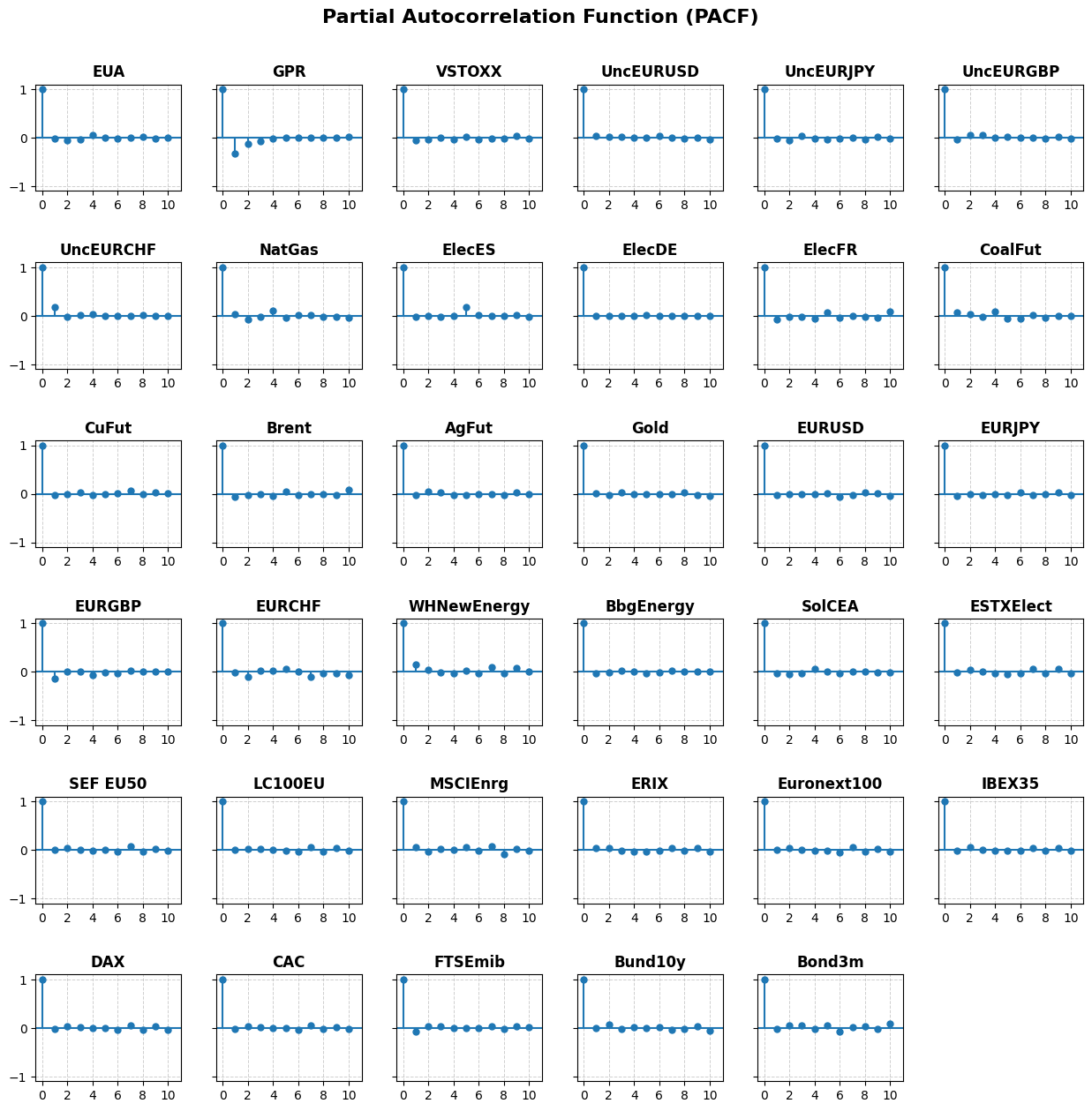}
        \caption{\textbf{Partial Autocorrelation Function (PACF) Plot of Financial Returns.} As in the ACF case, the PACFs decay to values close to zero from the first non-zero time lag.}
        \label{fig:pacf_rets}
        \end{figure*}
%
% End of appendix ---------------------------------------
\setcounter{section}{0}
\renewcommand{\thesection}{\arabic{section}}
\renewcommand{\thetable}{\arabic{table}}
\renewcommand{\thefigure}{\arabic{figure}}

\end{document}